\documentclass[prd,floatfix,twocolumn]{revtex4-1}
\usepackage{hyperref}

\usepackage{aas_macros}     
\usepackage{epsf} 
\usepackage{amssymb} 
\usepackage{amsmath}
\usepackage{graphicx}
\usepackage{subfigure}

\usepackage{color}

\DeclareMathAlphabet\mathbfcal{OMS}{cmsy}{b}{n}
\definecolor{darkgreen}{cmyk}{0.85,0.2,1.00,0.2} 
\definecolor{purple}{cmyk}{0.5,1.0,0,0}

\definecolor{orange}{rgb}{1.0,0.6,0.3}

\def\be{\begin{equation}}
\def\ee{\end{equation}}
\def\bea{\begin{eqnarray}}
\def\eea{\end{eqnarray}}

\def\lsim{\mathrel{\mathstrut\smash{\ooalign{\raise2.5pt\hbox{$<$}\cr\lower2.5pt\hbox{$\sim$}}}}}
\def\gsim{\mathrel{\mathstrut\smash{\ooalign{\raise2.5pt\hbox{$>$}\cr\lower2.5pt\hbox{$\sim$}}}}}
\def\half{\frac{1}{2}}
\def\({\left(}
\def\){\right)}

\def\bwt{\begin{widetext}}
\def\ewt{\end{widetext}}
\begin{document}

\title{Simulations of Galileon modified gravity:\\ Clustering statistics in real and redshift
space}
\author{Mark Wyman$^{1,2}$}
\email{markwy@oddjob.uchicago.edu}
\author{Elise Jennings$^{2,3}$}
\author{Marcos Lima$^{4}$}
 \affiliation{
$^{1}$ Department of Astronomy and Astrophysics, University of Chicago, Chicago, IL 60637 \\
$^{2}$ The Kavli Institute for Cosmological Physics, University of Chicago, 5640 South Ellis Avenue, Chicago, IL 60637\\
$^{3}$ The Enrico Fermi Institute, University of Chicago, 5640 South Ellis Avenue, Chicago, IL 60637\\
$^{4}$ Departamento de F\'isica Matem\'atica, Instituto de F\'isica, Universidade de S\~ao Paulo, SP, Brazil}
\date{\today}
\begin{abstract}
We use N-body simulations to study the statistics of massive halos and redshift space distortions for theories
with a standard $\Lambda$CDM expansion history and a galileon-type scalar field.
The extra scalar field increases the gravitational force, leading to enhanced structure formation.
We compare our measurements of the real space matter power spectrum and halo properties with fitting formula for 
estimating these quantities analytically.  We find that a model
for power spectrum, halo mass-function and halo bias, derived from $\Lambda$CDM simulations can fit the results 
from our simulations of modified gravity when $\sigma_8$ is appropriately adjusted.  We also study the redshift space distortions 
in the two point correlation function measured from these simulations, finding a  difference  in the ratio of the redshift space to real space 
clustering amplitude relative to standard gravity on all scales. We find enhanced clustering on 
scales $r>10$ Mpc$/h$ and increased damping of the correlation function for scales $r<9$ Mpc$/h$.
The boost in the clustering on large scales due to the enhanced gravitational forces cannot be mimicked in a standard gravity model by 
simply changing $\sigma_8$. This result illustrates the usefulness of redshift space distortion measurements as a probe of 
modifications to General Relativity.
\end{abstract}
\maketitle

The confluence of wide and deep galaxy redshift surveys with modern computing power
have brought us to the brink of a new era for cosmology, with precision tests of gravity and cosmology
on length scales from today's horizon scale to the small length scales where
non-linear density perturbations dominate. Because galaxies and clusters are the tracers 
used to study gravity, understanding how they and their host dark matter halos form and evolve is crucial.
The existence of structure moves galaxies out of the Hubble flow, and understanding 
the redshift space distortions, which arise from galaxy peculiar velocities, is an important way 
to extract even more information from observations.

Over the past two decades, great strides have been made in 
understanding the relationship between cosmological parameters
and structure formation. Large computational simulations have been performed and used to
test and calibrate analytic approaches for understanding the formation 
of non-linear structures. Testing a theory of gravity that deviates from General Relativity (GR) requires
checking whether the methods and results of the past still apply in the new model, especially
beyond linear perturbation theory. While this is true even in relatively modest alterations to gravity,
such as quintessence dark energy models, it is particularly important when the new gravitational
 physics introduces a new ``dark sector" for gravitation that alters the gravitational force. Observations
within the Solar System are in agreement with GR to great precision (for a review, see e.g. \cite{lrr-2006-3}). 
Hence, any new gravitational degrees of freedom must be suppressed on Solar System scales. 
There are, generally speaking, two known ways for this to occur. 1)  The effective ``charge"
that responds to the new gravitational force is reduced by the ambient conditions in the 
Solar System, called chameleon  \cite{Mota:2003tc,Khoury:2003aq,Hu:2007nk, Starobinsky:2007hu}
 or symmetron screening \cite{Hinterbichler:2010es}. 2) The gradients that generate the new force are reduced through non-linear effects, which has come to be known as Vainshtein screening \cite{Vainshtein:1972sx,2004JHEP...06..059N}.
Scalar fields that exhibit Vainshtein screening are generally called `galileons' because their self-interactions
are determined by an internal Galilean symmetry \cite{Nicolis:2008in}. 
Because both of these screening mechanisms are themselves inherently non-linear, it is important
when studying them to include both density non-linearities, as discussed above, as well as the non-linear
structure of the modified gravity theory under consideration. This requires numerical simulations.
Numerical simulations of galileon scalar fields will be the focus of this work.

In addition to the intrinsic interest of studying what kinds of new gravitational strength scalar fields
can exist in nature, 
there has been an important theoretical advance in the past two years:
a non-linearly complete and ghost-free theory of massive gravity in four dimensions has been found
\cite{Gabadadze:2009ja,deRham:2009rm,deRham:2010kj}. Giving the graviton a mass adds a new length scale into the
theory, making it possible to modify gravitation on long length scales in a consistent way. 
This new length scale, $r_c \propto 1/m_g$ (where $m_g$ is the graviton's mass), is assumed
to be of the order of the Hubble radius today.  Solutions of this theory have been found that 
exhibit cosmological acceleration even in the absence of a cosmological constant
\cite{deRham:2010tw, D'Amico:2011jj,Gratia:2012wt}. For the purposes of the present work, 
however, we note that this theory can be simplified
  in a decoupling limit to a theory with a
$\Lambda$CDM background cosmology and an extra galileon-type 
scalar field \cite{deRham:2010tw, deRham:2011by,Nicolis:2008in}
that manifests Vainshtein screening \cite{Sbisa:2012zk}. 
This gives additional motivation to our study of the 
cosmological effects of galileons.
 
Galileon theories generally, and massive gravity  in particular, contain many of the attractive aspects
of higher-dimensional braneworld constructions, such as the Dvali-Gabadadze-Porrati model \cite{Dvali:2000hr}
and its descendants (e.g. \cite{deRham:2007xp, Afshordi:2008rd}). However, they do not require
any extra dimensions and, unlike the DGP theory, are free of ghost-like instabilities  \cite{Hassan:2011hr, Hassan:2011ea}. They also
have the phenomenological advantage that their expansion history is expected to be very similar to that of $\Lambda$CDM, whereas 
DGP possessed a term linear in $H$ in its Friedman equations that was difficult to reconcile with expansion history observations,
especially on the ``self-accelerating" branch (see e.g. \cite{Song:2006jk}).

The calculations we present in this work were begun in \cite{Khoury:2009tk, Wyman:2010jp} in a somewhat different context. 
Those works drew their inspiration from a phenomenological 
version of a class of gravity models that arise when there are infinite volume extra
dimensions, as occurs in the Dvali-Gabadadze-Porrati model and its generalizations. Those set-ups
are generally known as ``cascading gravity" models.  
In \cite{Khoury:2009tk, Wyman:2010jp}, we performed N-body simulations using that model and
characterized the non-linear power spectrum of the dark matter fluctuations. The models we considered then had
a standard expansion history, much like the massive gravity theory;
they also similarly increase the growth of structure on linear length scales  
while recovering GR within collapsed structures. Fortuitously, the phenomenological model studied 
in  \cite{Khoury:2009tk, Wyman:2010jp} carries over nearly unchanged to the generalized galileon and massive gravity 
set-ups that are our focus in the present work.
In brief, the results of \cite{Khoury:2009tk, Wyman:2010jp} were that semi-analytic linear perturbation theory described the
model very well on long length scales, but that on scales $\lesssim 10$ Mpc/$h$,
existing analytic methods for including non-linearities in the power spectrum 
failed badly. We also quantified, in \cite{Wyman:2010jp}, the imprint of stronger gravity
on bulk flows, which are larger in models with an extra gravitational force.

In the present work, we deepen our understanding of structure formation in galileon  
models with a standard $\Lambda$CDM expansion history by studying the power spectrum, halo mass function, 
halo bias and
the correlation function in redshift space measured from N-body simulations of this model.
Our approach will follow the same pattern as work done in \cite{Oyaizu:2008sr,Oyaizu:2008tb,Schmidt:2008tn} for halos and \cite{Jennings:2012pt,Li:2012by} for
redshift space distortions for the $f(R)$ model, the phenomenology
of which is in some ways similar to that of the models we study. Our study complements and extends similar, earlier 
N-body simulations performed for the 
 ``normal" branch of the DGP model in Ref. \cite{Schmidt:2009sv}. 
 
 In the same spirit as  \cite{Khoury:2009tk, Wyman:2010jp},  we do not attempt to specialize to any
particular cosmological solution for massive gravity or galileons (e.g. \cite{deRham:2011by,Appleby:2012ba}). 
Instead, we make a series of phenomenological assumptions designed to isolate the growth-history effects of the galileon field
from its possible modifications to the Universe's expansion history. These assumptions are:
\begin{itemize}
\item Assume exactly $\Lambda$CDM expansion history.
\item Compute the GR / Newtonian force as in $\Lambda$CDM.
\item Additionally solve the equations governing the galileon scalar field.
\item Assume the dynamical potential is the sum of the Newtonian and extra scalar field contributions. 
\end{itemize}
These assumptions mean our results will not be tied to any particular cosmological solution, but can be broadly applied to any model with galileon scalars that has
an expansion history close to $\Lambda$CDM. On the other hand, nothing we find can be definitively associated
with, for instance, the model of massive gravity {\it per se}, so our results will need to be revisited as data and theoretical understanding
of the model improve.

Our chief results are these.
\begin{enumerate}
\item The halo mass function and the linear halo bias at $z=0$ are altered by the increased gravitational force in a way degenerate with an increase in $\sigma_8$ in the standard $\Lambda$CDM model. This is in contrast with chameleon / $f(R)$ theories, which generate a different modification to the halo mass function (e.g. \cite{Li:2011uw}). However, since the apparent $z=0$ normalization of this model is itself a function of redshift, measurements of the halo mass function at different redshifts would break this degeneracy. Similar conclusions hold for the real space power spectrum on linear and mildly non-linear 
scales.
\item Large-scale redshift space correlations are enhanced in these models in a way that cannot be mimicked within the $\Lambda$CDM paradigm, because velocity space alterations directly probe the increased long range gravitational force generated by the galileon field. In chameleon / $f(R)$ theories, which do not have very long range force modifications, this effect is absent \cite{Jennings:2012pt}.
\end{enumerate}

We note that the numerical approach we use is improved as compared with the N-body simulations described in \cite{Khoury:2009tk}. 
In particular, we have replaced the phenomenological approximation used there with a more
sophisticated multigrid algorithm for solving the nonlinear equation of motion of the 
extra scalar field that generates much of the model's new physics. See Appendix \ref{numerics} for more
details. Results from this 
improved code first appeared in \cite{COSMO09, Wyman:2010jp}.
Nonetheless, the results of this much more computationally costly 
approach are in surprisingly good agreement with those in our previous work.

For our numerical simulations,we will take two particular values for the model parameter $r_c$, which in the context 
of massive gravity represents the graviton's Compton wavelength: $r_c=1089$ Mpc and $r_c=1665$ Mpc.
These values of $r_c$ were chosen based on the results from \cite{Wyman:2010jp} (see also Fig.~\ref{fig:lineartheory}), which found (normalizing to the
CMB) that $r_c=1089$ Mpc would have
 a linear power spectrum at $z=0$ with $\sigma_8=0.92$, in conflict with current data, while
$r_c=1665$ Mpc would have a linear power spectrum at $z=0$ with $\sigma_8=0.88$, which is on the borderline
of being ruled out by current data.

For our cosmology, we assume a spatially flat Universe with
$\Omega_{M} = 0.24 \equiv \Omega_M^0$, $\Omega_\Lambda = 0.76$, and
$h=0.73$ (N.B., after equation 11, the notation $\Omega_M^0$ is used for 
the present value of $\Omega_M$).
Wherever we do not otherwise specify,
we choose an initial amplitude for fluctuations that would generate
$\sigma_8=0.8$ in the usual GR context. We also take the spectral tilt
$n_s=0.96$. Finally, wherever we do not write dimensionful
constants explicitly, we will use units for which $\hbar = c = M_{\rm pl} = 1$. 

\section{Galileons and Massive Gravity}
\label{review}

There are few ways to modify gravity that are not ruled out by Solar System tests. 
The simplest way to alter gravity phenomenologically is to add a new scalar field with a gravitational-strength coupling, in the spirit of the Brans-Dicke model. 
However, it is important also to look for modifications that have more sophisticated theoretical motivation.  Since the chief motivation 
for modifying gravity is dark energy, which only began accelerating the Universe's expansion rate at late times, we would like to find
modifications to gravity that only appear at long length scales. 

Because of the equivalence principle, GR on its own cannot exhibit new behavior at long length scales. This is related
to the fact that the graviton is massless, and hence has no built-in length scales other than the Planck scale. However, if the
gravitational force were carried by a massive particle, the equivalence principle would no longer be in effect.
In 1939, Fierz and Pauli demonstrated that a massive graviton can be defined perturbatively \cite{Fierz:1939ix}. The simple
Fierz-Pauli model was found to be inconsistent with observations (even in the limit when the graviton mass is taken to zero) 
because of the so-called vDVZ (van Dam, Veltman, Zakharov)
discontinuity \cite{vanDam:1970vg,Zakharov:1970cc}. This ``discontinuity" arises from the physical fact that a massive spin-2 graviton
has more degrees of freedom than the massless one. In practice, it implies a disagreement between Newton's constant  measured
by gravitational lensing as compared with the gravitational force on massive particles. Vainshtein demonstrated that non-linear
completions of the Fierz-Pauli evade the discontinuity because the extra degrees of freedom have such strongly non-linear self-interactions
that they decouple  from everything else either near matter sources or as the graviton mass goes to zero  \cite{Vainshtein:1972sx}, but Boulware and Deser showed
that generic non-linear completions introduce an extra propagating mode that is a ``ghost" (a particle with negative kinetic energy), implying
that such theories are internally inconsistent \cite{Boulware:1972zf}. However, recent work has shown that not every non-linear completion has a ghost. In particular, de Rham, Gabadadze, and Tolley (dRGT) found a class of completions  \cite{Gabadadze:2009ja,deRham:2009rm,deRham:2010kj} that were shown to be ghost free to all orders of perturbation theory \cite{Hassan:2011hr,Hassan:2011ea}. As expected, the
resulting theory of massive gravity propagates extra degrees of freedom beyond those present in GR. In particular, the graviton
now has two vector components and one scalar component in addition to GR's two tensor parts. 

Although the resulting dRGT theory is both complicated and highly non-linear, at the phenomenological level we can
gain insight by studying the theory in what is known as a decoupling limit. In the decoupling limit, we assume that the
mixing among the different components of the graviton is small. The practical upshot of this is that the theory reduces to a scalar-tensor theory,
albeit one in which the scalar field is a \lq galileon\rq \,, which we define below \cite{deRham:2010tw, deRham:2011by,Nicolis:2008in}. 
In this simpler theory, we can easily see the origin of the vDVZ discontinuity \cite{vanDam:1970vg,Zakharov:1970cc} and its resolution. 
Assuming linear theory and for wavelengths small compared with the horizon scale (where the decoupling limit is valid), the
effective Poisson equation for the  potential, $\Psi_{\rm dyn}$, felt by massive particles becomes 
\be
k^2 \Psi_{\rm dyn} = -4\pi G \left(1 + \frac{1}{3}\right) \rho \,.
\label{poissonflat}
\ee
The extra $1/3$ here is the manifestation of the additional scalar field's force. If this linear equation were exact, this theory
would be ruled out by Solar System tests (e.g. \cite{lrr-2006-3}).  However, the equation of motion for the extra scalar field, which we shall
call $\varphi$, is strongly modified by non-linear effects. Making the standard and well-motivated assumption that the scalar's time
derivatives are small compared with its spatial derivatives on sub-horizon scales, the approximate equation of motion for $\varphi$ (near flat space, i.e., not
in an FRW-like cosmological setting) has the form \cite{deRham:2010tw}
\begin{align}\label{pieqn}
\nabla^2 \varphi & + \frac{r_c^2}{3} [ (\nabla^2\varphi)^2
- (\nabla_i\nabla_j\varphi)(\nabla^i\nabla^j\varphi) ]   + \cdots = \frac{8\pi\,G }{3} T,
\end{align}
where again, in the context of massive gravity, the model parameter, $r_c \equiv (\hbar/c ) /m_g$, $m_g$ is the mass of the graviton, 
and $r_c$ is its associated Compton wavelength. $T$ is the trace of the stress-energy tensor. The theory generally has two more free parameters beyond $r_c$ that
 would give additional contributions in this limit. However, they will not be used in our study, so we simply inserted an ellipsis in the equation
 above to represent these further complications. Scalars with these kinds of equations (both the simple form that appears in Eqn. \ref{pieqn} 
 as well as its generalization)
 are called `galileons' because
 their equations exhibit an analog of Galilean invariance: their dynamics are left the same under the replacement $\varphi \to \varphi + c  + b_\mu x^\mu$;
 this is seen at the level of the equations of motion by the fact that the field is always differentiated twice. 
Eq.~\ref{pieqn} is highly non-linear in the gradients of the scalar field, but can be solved exactly in the cases of spherical 
symmetry and planar symmetry. A planar ansatz sets all of the non-linear terms to zero, whereas in the spherically symmetric
case we find the expected suppression of the extra force for large densities. This suppression sets in at a new characteristic
length scale $r_*$, which is known as the Vainshtein radius; in the point mass case, it may be defined as
\be
r_* = (2GM r_c^2)^{1/3} =  ( r_s r_c^2)^{1/3}.
\ee
Roughly speaking, the extra force is unscreened at distances larger than $r_*$ and is suppressed at distances smaller
than $r_*$. 

\subsection{Phenomenological model}\label{pheno}

As mentioned in the Introduction, for our numerical study we will make some simplifying phenomenological assumptions.  The first of these is that we will assume an exactly $\Lambda$CDM expansion history. We do this for two reasons.
First, we want to isolate the effect of the extra scalar field on the growth history for comparison with standard gravity.
It is thus useful to keep the expansion history fixed to avoid confounding effects. Secondly, currently-known cosmological
solutions for galileons and massive gravity are very close (or identical) in their expansion history to $\Lambda$CDM,
even in the absence of a cosmological constant. 
Now, in addition to an expansion history, we must also assume some form of cosmological screening for the extra scalar field.
That is, we expect the extra scalar force to be screened when the Universe's horizon is smaller than its own ``Vainshtein" radius,
and for the strength of the extra scalar force gradually to increase as the horizon scale grows. 
That is, the maximum strength of the force at a given time will be
\be
\partial \varphi_{\rm max} = \frac{1}{3 B(a)} = g_{\rm linear \, theory},
\ee
where $B(a)$ is a function that depends on the cosmological evolution of the background.
Note that this factor of $1/3$ is the same as the one that appears in Eq.~\ref{poissonflat}; when $B=1$, we recover
a scalar force whose strength is exactly $1/3$ that of GR -- the maximum strength this force can achieve. This
combination appears as the function $g$ in the linearized theory of this model, and is plotted in the lower panel of Fig.~\ref{fig:lineartheory}.
The function $B(a)$ could be solved for directly in the DGP model. For the present study, we will follow our previous work \cite{Khoury:2009tk,Wyman:2010jp} and adopt a form inspired by the DGP model:
\be
B(a) \equiv 1 + 2 \left(Hr_c\right)^{2}\left(1+\frac{\dot{H}}{3H^2}\right).
\ee 
The change in this function relative to the DGP model is that our function is controlled by $(H r_c)^2$, whereas the
function in the DGP case went as $H r_c$. This alteration represents a more rapid turn-on of the scalar force relative to DGP.
Although this dependence was heuristically anticipated in our previous works, recent more rigorous attempts to study perturbations
in massive gravity have found a function with the expected $(H r_c)^2$ dependence ( Eq. (48) in \cite{deRham:2011by}).

Another simplifying phenomenological assumption is
that we will drop the higher-order pieces of Eq.~\ref{pieqn}. These generally have the same form
as the first piece, but are raised to higher powers.
 We do this mainly to make the numerical computations tractable, but we do not expect that it will make
an appreciable change in the large scale dynamics of the theory. We can argue for this as follows.
In spherical symmetry, we can solve the equation exactly, with and without the extra terms (see \cite{Nicolis:2008in}). In both cases,
the characteristic radius that appears in the resulting solution has the same scaling with the mass of the source
and $r_c$, the graviton's Compton wavelength, i.e. $r_* \propto (M r_c^2)^{1/3}$. Hence there is no qualitatively new
behavior or new length-scale introduced when the higher-order terms are included. The solutions do 
have different behavior as $r \rightarrow 0$ (see e.g. \cite{Wyman:2011mp}), but our simulations will not be able to resolve
the length-scales on which these differences manifest themselves. 

We will thus solve the following system of equations:
\begin{align}\label{sys}
\nabla^2 \Phi_N & = 4 \pi G \delta \rho \, , \\
\nabla^2 \varphi  + \frac{r_c^2}{3 B} [ (\nabla^2\varphi)^2
- (\nabla_i\nabla_j\varphi)(\nabla^i\nabla^j\varphi) ]  &= \frac{8\pi\,G }{3 B} \delta \rho, \nonumber
\end{align}
where $\Phi_N$ is the usual Newtonian potential.  Note that we have assumed $\delta \rho$, rather than $\rho$ is the source
for $\varphi$. This is an assumption that the phenomenological $\varphi$ field we are studying is sourced
by local overdensities, and that any global solutions of $\varphi$ have been absorbed into the $\Lambda$CDM-like
background expansion.
These equations result in two gravitational potentials, $\Phi_N$ and $\varphi$, which are combined into a single dynamical
potential for moving particles:
\be
\Psi_{\rm dyn} = \Phi_N + \half \varphi.
\ee
This is the generalization of the $\Psi_{\rm dyn}$ that appears in Eq.~\ref{poissonflat}. 

\subsection{Summary of simulations}
The results reported here were compiled from a large number of computational runs with 512$^3$ particles on a 512$^3$ grid. 
Each set of runs was comprised of simulations from $z=49$ 
to $z=0$ (in fixed $\Delta a = 0.0025$ steps) performed in 4 boxes of sizes $L_{\rm box}=64, 128, 256$, and 400 Mpc$/h$, with three different gravitational theories: ordinary Newtonian gravity, plus two modified gravity runs assuming inverse graviton mass $r_c = 1089$ and $1665$ Mpc. 
  For each set, a different initialization seed was used to set initial conditions, then that set of initial conditions was used for each of the three different gravity theories to minimize inter-run variance. We additionally employed a technique used in \cite{Stabenau:2006td}
that normalizes the overall amount of initial inhomogeneity for each initialization seed. Although these choices reduce
our ability to directly compare our results to data, they greatly assist our comparisons with standard gravity, which is our focus. Some more computational
details are discussed in Appendix \ref{numerics}.
Our suite of simulations is as follows:
\begin{itemize}
\item 8 sets of runs with outputs at only $z=0$ for all three models, initialized with $\sigma_8({\rm GR})=0.8$
\item 2 sets of runs with outputs at $z=0.8,0.6,0.4,0.2$ and $0$ for all three models, initialized with $\sigma_8({\rm GR})=0.8$.
\item 1 set of GR-only runs with  outputs at $z=0.8,0.6,0.4,0.2$ and $0$, initialized with $\sigma_8({\rm GR})=0.88$ and $\sigma_8({\rm GR}) =0.92$.
\item 1 set of runs with the non-linearities in Eqs.~\ref{sys} turned off (i.e. $r_c=0$) for the two galileon models.
\end{itemize}
The initial conditions for the runs were set, as in \cite{Klypin:1997sk}, using the Zeldovich approximation to displace particles. See \cite{Klypin:1997sk} for more details.
We note that our simulations methods are very similar to those used in \cite{Schmidt:2009sg, Schmidt:2009sv}. 

\section{Structure Growth: Linear Theory}

\begin{figure}[htbp] 
   \centering
   \includegraphics[width=0.45 \textwidth]{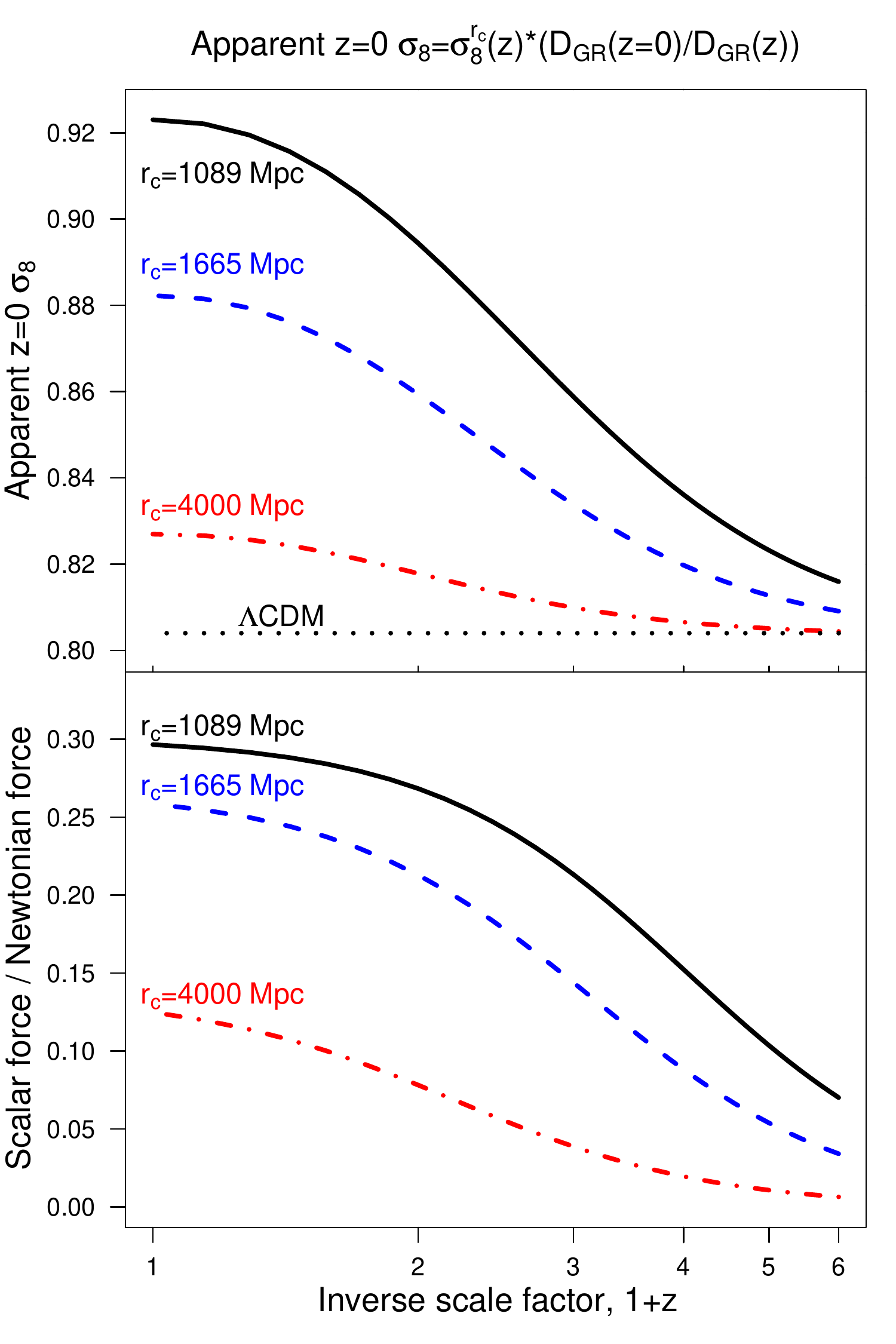} 
   \caption{ Results from linear theory. In the top panel, we show the $z=0$ amplitude of the dark matter linear power spectrum that would be inferred by standard
   methods for different values of  $r_c$. This measure, Apparent $\sigma_8$, is defined in the text and in the plot title where $D_{\tiny \rm GR}$ is the growth factor in standard gravity. In the lower panel,
   we plot the ratio of the strength (in linear theory) of the extra scalar force as compared with the Newtonian force; this appears as $g = 1/(3 B(a))$ in the text. Although the extra force has in principle a maximum strength of $1/3$ the Newtonian force, the phenomenological model we adopt suppresses this under the assumption that the background density of space will modulate the strength of the extra scalar's force, as was found in the DGP model. }
   \label{fig:lineartheory}
\end{figure}

We can linearize the equations given in Eqs.~\ref{sys} to give a simple view of how linear structure
formation is altered in this model. In Fourier space, we can write the evolution of the overdensity
mode $\delta=\delta_k$ as 
\be
\delta''+ \(2+\frac{H'}{H}\) \delta'  =  -  \frac{k^2}{a^2 H^2} \Psi_{\rm dyn}\, , 
\ee
where $'$ indicates the derivative with respect to e-folding time, $d \ln a$. For linear theory, we can use
 \be
\left(\frac{k^2}{a^2} \right)\Psi_{\rm dyn} = -4\pi G (1+g)  \bar{\rho}  \delta_k,
\ee 
with $g$ is defined by
\be
g = \frac{1}{3}\cdot\frac{1}{1 + 2 \left(Hr_c\right)^2\left(1+ \frac{{H'}}{3H} \right)}.
\label{g}
\ee
We plot $g$ as a function of redshift in our particular cosmology in Fig.~\ref{fig:lineartheory}.
Using these relationships, we can find the combined equation
\be
\label{linthy}
\delta''+ \(2+\frac{H'}{H}\) \delta' = \frac{3}{2} \Omega_M(a) \( 1+ g\) \delta
\ee
where, for easy reference,
$$
 \Omega_M(a) = \frac{\Omega_M^0}{a^3} \frac{H_0^2}{H^2}.
$$
For a cosmology with only matter and $\Lambda$, we have 
\be
\frac{H'}{H} = - \frac{3}{2} \frac{\Omega_M^0 a^{-3}}{\Omega_M^0 a^{-3} + (1-\Omega_M^0)}.
\ee

For our linear theory solutions, we take initial conditions from CAMB \cite{Lewis:1999bs} at $z=50$. At this redshift,
the extra scalar has no effect on cosmology, so we are justified in using a standard gravity-based code for generating
our initial conditions. We then
use Eq.~\ref{linthy} to numerically evolve the Fourier density modes from $z=50$ to $z=0$ for different
values of $r_c$. We plot the results in Fig.~\ref{fig:lineartheory}.

The results plotted in Fig.~\ref{fig:lineartheory} are presented in a somewhat unusual format, which we will explain.
As we have stated before, the extra scalar force only starts to influence growth history at relatively late times, and then
grows in its influence as time goes on. Hence, it is useful to quantify the deviation of the power spectrum generated including
the extra scalar from the power spectrum that would have been generated by GR. We do this by means
of a measure we call Apparent $z=0$ $\sigma_8$. Recall that $\sigma_8$ is a commonly used quantity that encodes
the normalization of the power spectrum by measuring the
matter fluctuation within 8 Mpc$/h$ spheres. In general, we can define the variance
of the linear density field within spheres of radius $R$, $\sigma_R$ via
\be
\label{sigR}
\sigma_R^2 \equiv \frac{1}{2\pi^2} \int_0^\infty k^2 \, P(k,z=0) W^2_R(k) {\rm d}k\,.
\ee
To get $\sigma_8^2$, we take $W_8(k) = 3j_1(kR_8)/kR_8$, with $R_8 = 8$  Mpc$/h$ and $j_1$ a spherical Bessel function. 

Observational cosmology involves making measurements of the power spectrum of density
perturbations at many redshifts. For comparison purposes, these measurements are commonly extrapolated
to $z=0$ using linear theory and reported as measurements of $\sigma_8$. It is in this sense that the amplitude
of the CMB power spectrum gives a measurement of $\sigma_8$. In Fig.~\ref{fig:lineartheory}, we make use of a
similar algorithm:
\begin{itemize}
\item For a set of $z_j$ between 5 and 0, we evolve the power spectrum from $z_{\rm init}=50$ to $z=z_j$ using the modified gravity equations, Eq.~\ref{linthy}
\item Beginning at $z=z_j$, we take the modified gravity-generated power spectrum and then evolve it to $z=0$ using 
the growth factor, $D_{ \tiny \rm GR}$, in standard gravity,
essentially setting $g=0$ in Eq.~\ref{linthy}.
\item We evaluate $\sigma_8$ using the power spectrum generated by this procedure, and record it as the Apparent $\sigma_8(z_j)$. 
\end{itemize}

These linear theory results give a simple qualitative picture of how this model modifies the growth of structure:
\begin{enumerate}
\item Growth is the same as in GR when the Universe's horizon scale is smaller than its Vainshtein radius.
\item As the Universe expands, its horizon grows larger than its Vainshtein radius, allowing the extra scalar to begin to accelerate the growth of structure (see
the second panel of Fig.~\ref{fig:lineartheory}.
\item As the extra scalar force operates, structure grows faster than it would in GR, leading to more large scale structure and a larger Apparent $\sigma_8$.
\end{enumerate}

\section{Results: Halo Properties}
\label{sec:halos}

In this section we study the halo abundance and linear bias, as inferred from 
simulations, and how standard fitting prescriptions calibrated in $\Lambda$CDM 
simulations can match the galileon modified gravity simulated results. 
The techniques we employ to extract the halo catalog, 
mass-function and bias are the same as those presented in \cite{Schmidt:2008tn} in the context 
of $f(R)$ models. 

\subsection{Halo Abundance}

\begin{figure*}[ht!] 
   \centering
   \includegraphics[width=0.45 \textwidth]{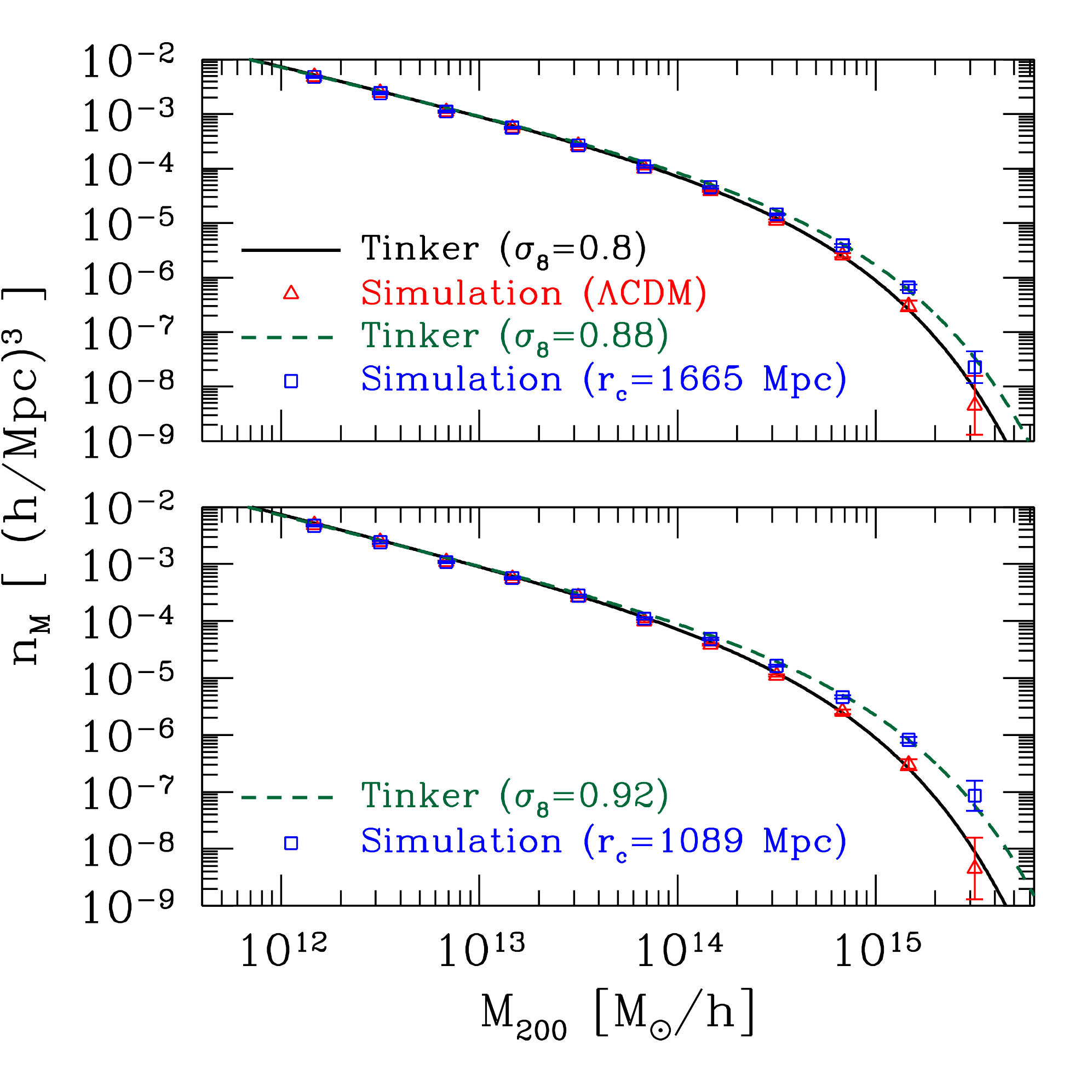} 
   \includegraphics[width=0.45 \textwidth]{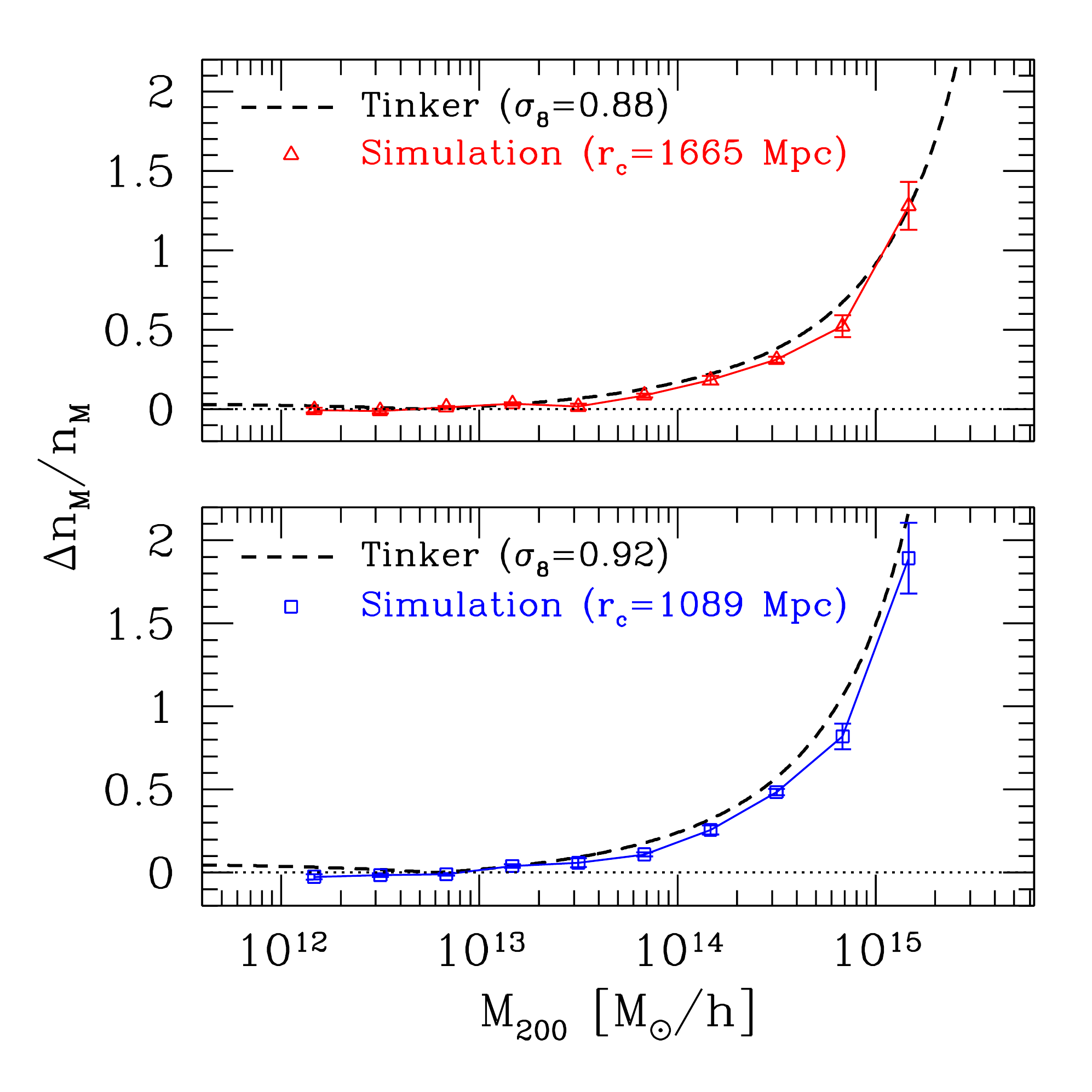} 
   \caption{
   (Left): Halo mass function at $z=0$ measured from   $\Lambda$CDM (red triangles) and galileon modified gravity simulations 
(blue squares) with $r_c=1665$ Mpc (upper panel) and $1089$ Mpc (lower panel). The  $\Lambda$CDM predictions of 
   Tinker et al. \cite{Tinker:2008ff} for $\sigma_8 =0.8$ (solid curves) and $\sigma_8=0.88,0.92$ (dashed curves) are also plotted. 
  (Right): Percent difference for the mass function measured from modified gravity simulations relative to a 
$\Lambda$CDM simulation with $\sigma_8=0.8$ (triangles and squares), and Tinker predictions relative to $\sigma_8=0.8$ (dashed lines). 
}
   \label{fig:mf}
\end{figure*}

We detect halos using a spherical overdensity 
(SO) halo finder, and define halo masses within an overdensity $\Delta=200$ 
with respect to the mean background density $\bar{\rho}_m$.  
The halo mass function is obtained dividing the number of halos in a given mass bin by the 
comoving volume of the simulation box and the mass bin size. 
To be conservative, we keep only halos with more than 6400 particles in each box and combine the mass functions 
from all boxes and runs at $z=0$. 

We compare our measured mass function with the fitting formula of Tinker et al. \cite{Tinker:2008ff}, 
obtained from high-resolution $\Lambda$CDM simulations, and given by
\bea
n_M= \frac{dn(M,z)}{d \log M}=f(\sigma)\frac{\bar{\rho}_m}{M} \frac{d \ln \sigma^{-1}}{d \log M}
\eea
where $\sigma(M)=\sigma_R$ is the variance of the linear density field 
for a mass $M=4\pi R^3 \bar{\rho}_m/3$ contained in a 
sphere of radius $R$ at the mean background density (see Eqn. \ref{sigR}),
and
\bea
f(\sigma)=A\left[\left(\frac{\sigma}{b}\right)^{-a} + 1\right] e^{-c/\sigma^2}.
\eea
For $\Delta=200$ we set parameter values $A=0.186$, $a=1.47$, $b=2.57$ and $c=1.19$.

In Fig.~\ref{fig:mf}, we show the mass function as a function of halo mass for simulations
and predictions. In the panels on the left, the simulation-derived mass function is plotted  
as triangles for the $\Lambda$CDM simulations and squares for galileon modified gravity simulations 
with $r_c=1665$~Mpc (top) and $r_c=1089$~Mpc (bottom). Mean values and error bars are derived 
from volume-weighted bootstrap samples in order to reduce sample variance, similarly to  \cite{Oyaizu:2008tb, Schmidt:2008tn}. 
In the right hand panels, we show the relative change $\Delta n_M/n_M=(n_M^{r_c}/n_M^{\Lambda \text{CDM}}-1)$ measured from the 
$r_c=1665$~Mpc (triangles) and $r_c=1089$~Mpc (squares) simulations.  
Here we have neglected the last measured point shown on the left panels, corresponding to the most massive halos, 
 because the number of halos in this bin is of order unity, making any measure of relative differences meaningless.  
We also show the predicted percent difference between the Tinker et al. \cite{Tinker:2008ff} fit 
for $\Lambda$CDM with $\sigma_8=0.88$ and $0.92$ relative to that for $\sigma_8=0.8$.

We find that the mass function in the modified gravity scenario can be fit with a 
$\Lambda$CDM mass function 
with higher $\sigma_8$.  
This degeneracy 
 prevents  low redshift cluster abundance measurements from being able to distinguish galileon modified gravity from $\Lambda$CDM.
 In principle this degeneracy can be broken by cluster observations at higher redshifts. 
 As we go back in time, the extra force is weaker and the cluster measurements tend to agree with those of 
 a true $\Lambda$CDM universe (in our case with $\sigma_8=0.8$). 
 From Fig.~\ref{fig:lineartheory} we see for instance that the apparent $\sigma_8$ for the $r_c=1089$ Mpc 
 case changes from $0.92$ at $z=0$ to $0.86$ at $z=2$. For higher values of $r_c$, this change is smaller and it becomes 
 harder to distinguish galileons from $\Lambda$CDM.

\begin{figure*}[htbp] 
   \centering
      \includegraphics[width=0.45 \textwidth]{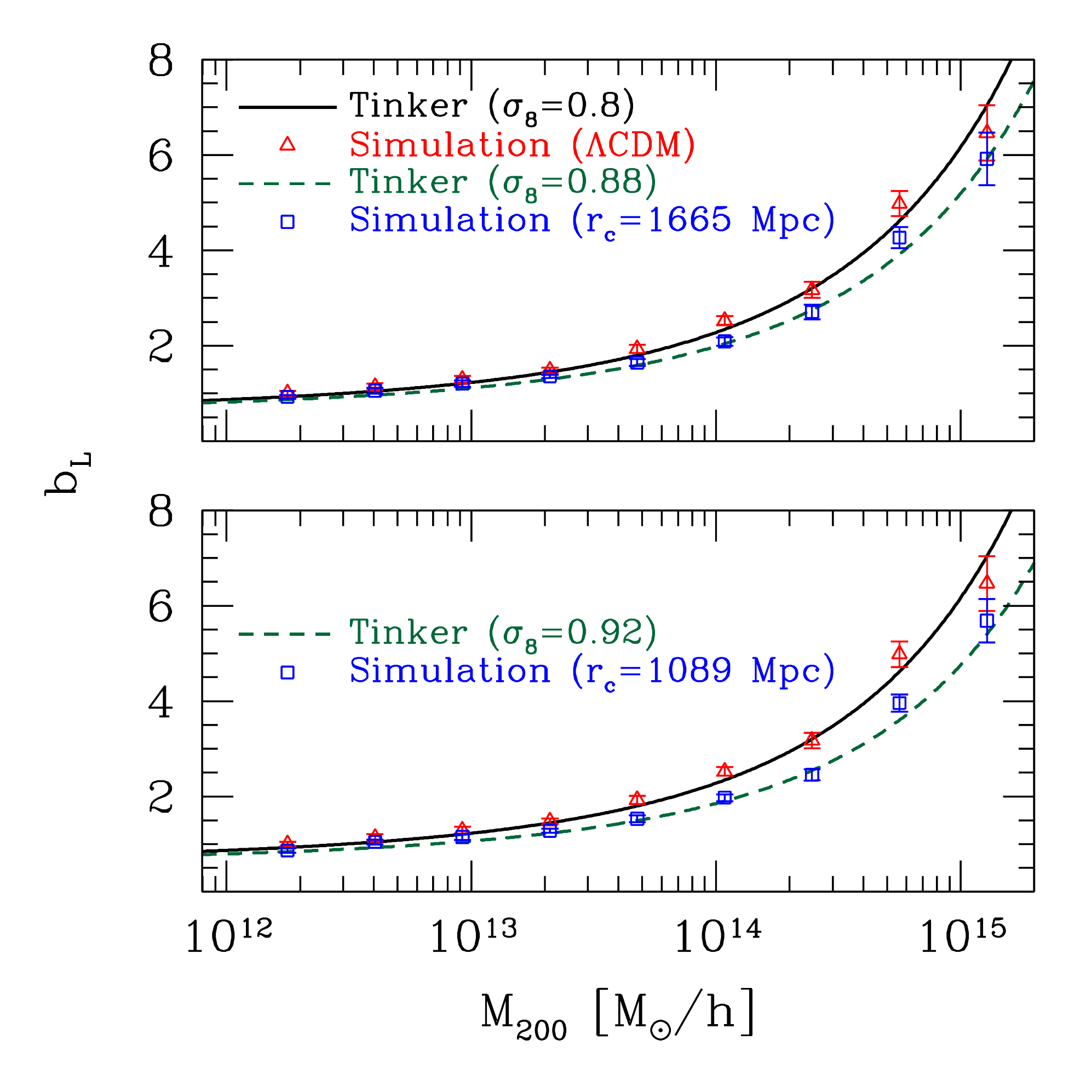} 
      \includegraphics[width=0.45 \textwidth]{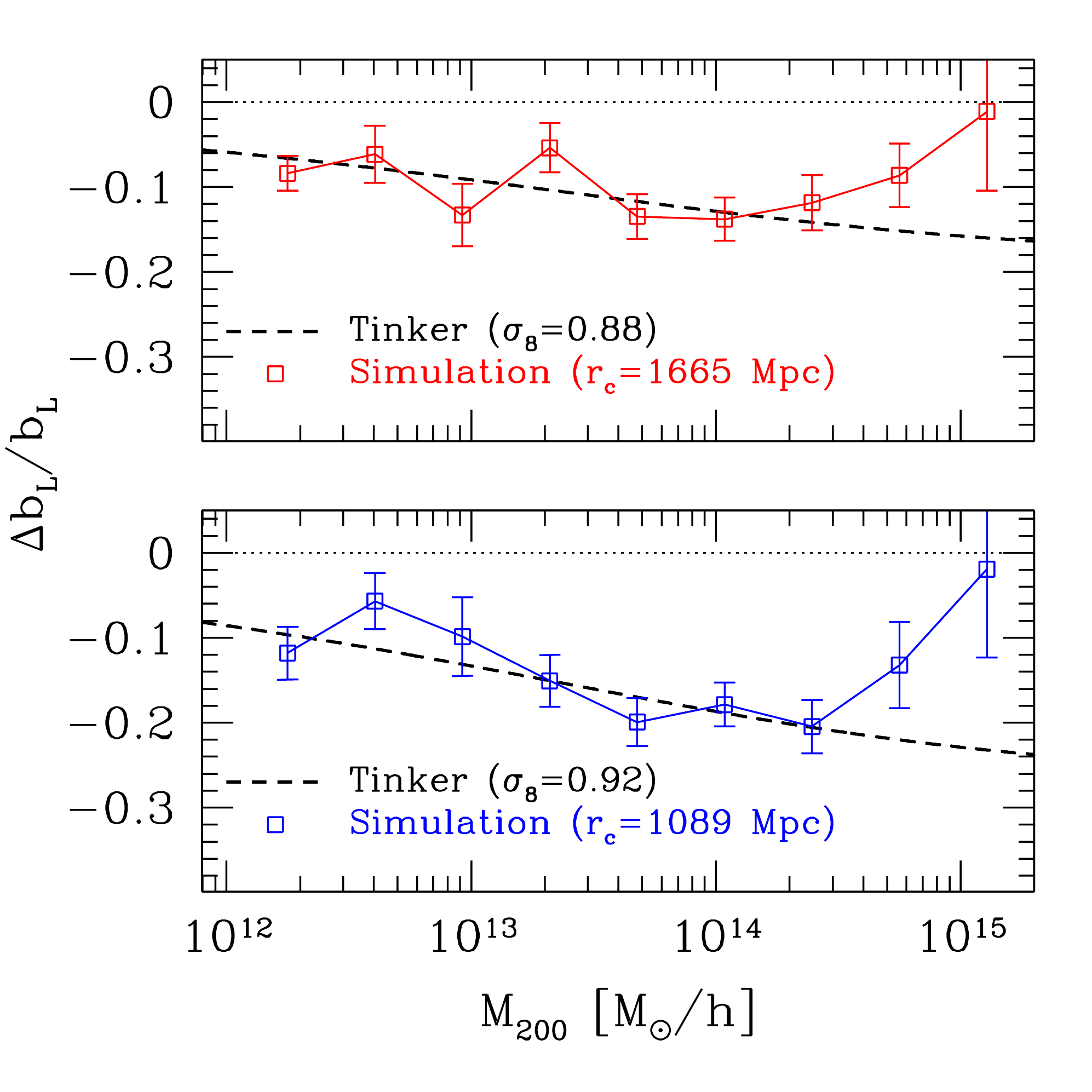} 
   \caption{Linear halo bias at $z=0$ for the same cases shown in Fig.~\ref{fig:mf}. 
   The results for the halo bias are noisier than those for the mass function, but similarly display the degeneracy between 
   galileon modified gravity and  $\Lambda$CDM with larger $\sigma_8$ at single redshifts.}
   \label{fig:bias}
\end{figure*}

\subsection{Halo Bias}

Halo bias characterizes the clustering of galaxies with respect to the underlying mass distribution, 
and is assumed to be scale independent on large scales which are still in the linear regime. 
For halos of a given mass $M$, we compute the halo bias from the simulations using the definition 
$b(k,M)=P_{hm}(k)/P_{mm}(k)$, where $P_{mm}(k)$ is the dark matter power spectrum 
and $P_{hm}(k)$ is the halo-mass cross spectrum. This choice allows us to partially reduce 
the shot noise that would result if we used the halo-halo auto-spectrum $P_{hh}(k)$.
To obtain the linear halo bias, we fit a polynomial to  $b(k,M)$ using the first 10 
values of $k$ and extrapolate to the lowest $k$ value, i.e. $b_{L}(M)=b(k=k_{\rm min},M)$.

Similarly to the mass-function results, we then compare our measured bias with the fitting 
formula of Tinker et al.\cite{Tinker:2010my}, calibrated from high-resolution $\Lambda$CDM simulations as
\bea
b_L(M)
         &=& 1 - A\frac{\nu^a}{\nu^a+\delta_c^2} + B \nu^b + C \nu^c,
\eea
where $\nu(M)=\delta_c/\sigma(M)$ and $\delta_c=1.686$.
We fix values for parameters $A$, $a$, $B$, $b$, $C$ and $c$ 
appropriate for our SO mass definition of $\Delta=200$ \cite{Tinker:2010my}. 

In Fig.~\ref{fig:bias}, we show the linear halo bias for the same cases shown 
in Fig.~\ref{fig:mf}.  The bias measurements are noisier than those of 
the mass-function, mainly due to the shot noise in the halo-mass 
cross-spectrum.  The percent difference $\Delta b_L/b_L=(b_L^{r_c}/b_L^{\Lambda \text{CDM}}-1)$ 
is less pronounced than the corresponding differences in the abundance. 
Nonetheless, it is consistent with a $\Lambda$CDM model with a higher value of $\sigma_8$. 

Altogether, the halo properties at low $z$ show that it should be hard to break 
the degeneracy between galileon modified gravity and $\Lambda$CDM models with larger 
values of $\sigma_8$ unless one can make measurements at various
redshifts. Even though it is beyond the scope of this work, it should be possible 
to develop improved methods that break universality in order to include the effects 
of modified gravity more accurately in the halo properties. In the remaining of the 
paper we investigate whether a more unique signature of galileon models 
can be seen in the two-point statistics of galaxies and in their redshift-space distortions.

\section{Results: Clustering in real and redshift space}
In this section we present the power spectrum measured from the simulations 
in real space (Section \ref{subsection:real_pk}) and  review the linear perturbation theory of redshift space
distortions (Section \ref{subsection:rsd_theory}). The method used to estimate the correlation function in real
and redshift space is outlined in Appendix \ref{section:xi}. Our results showing the redshift space clustering signal 
in galileon models  compared to $\Lambda$CDM are presented in Section \ref{subsection:rsd_results}.

\subsection{Power spectra in real space \label{subsection:real_pk}}

\begin{figure*}[htbp] 
   \centering
      \includegraphics[width=0.45 \textwidth]{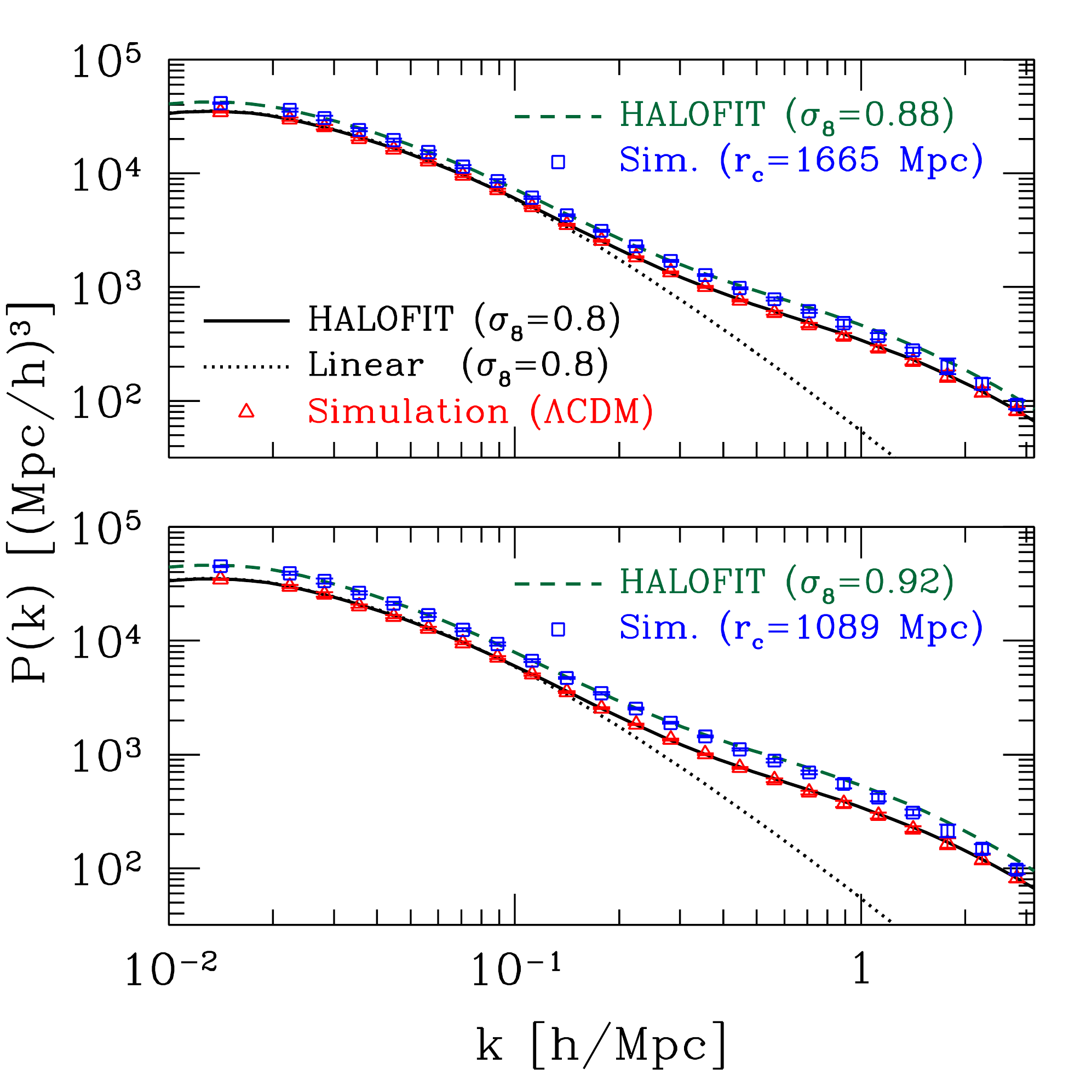} 
      \includegraphics[width=0.45 \textwidth]{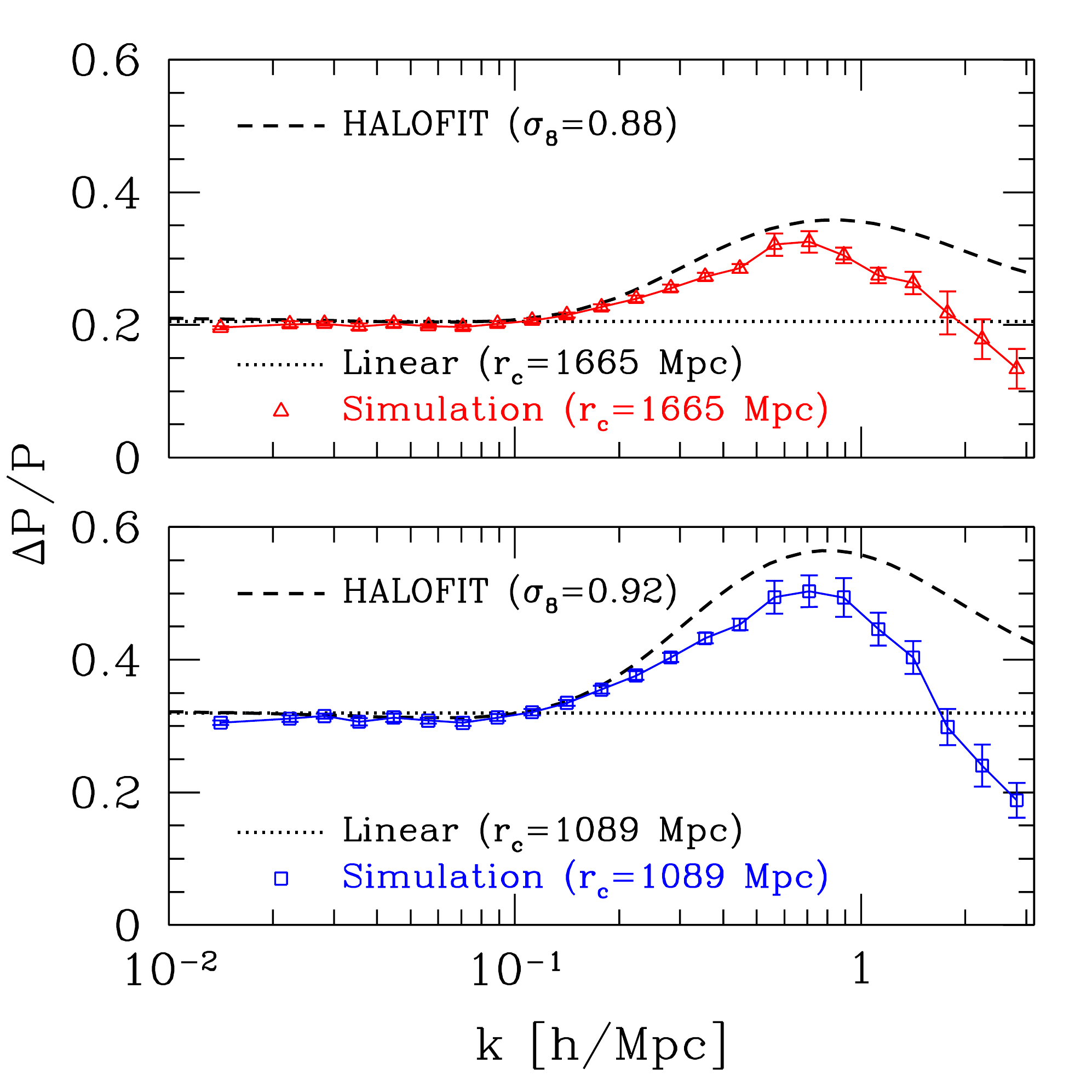} 
   \caption{Power spectrum at $z=0$ for the same cases shown in Figs.~\ref{fig:mf} and \ref{fig:bias}. The left panels 
   show the simulated power spectra for the $\Lambda$CDM simulation (triangles) and for the galileon simulations 
   (squares) with $r_c=1665$ Mpc (top) and $r_c=1089$ Mpc (bottom). Also shown are the linear power spectrum 
   for $\Lambda$CDM with $\sigma_8=0.8$ and the HALOFIT power spectra for $\sigma_8=0.8$ (solid line) 
   0.88 (dashed top) and 0.92 (dashed bottom). The right panels show relative deviations from these cases. 
   On deeply non-linear scales the Vainshtein screening is effective and it is not possible to fit the galileon 
   simulations using standard $\Lambda$CDM fits with higher values of $\sigma_8$. }
   \label{fig:ps}
\end{figure*}

In order to estimate the power spectra, for each box of a given size and cosmological model, 
we define a density field in the grid of $512^3$ points using the Cloud-In-Cell (CIC) method. 
The density field is Fourier transformed and the power spectrum of each box is estimated averaging 
the band-power in each Fourier mode bin. We then 
combine the various boxes using the same volume-weighted bootstrap averaging procedure 
used for the halo mass-function and bias.

In Fig.~\ref{fig:ps} we show the power spectra for the same cases displayed in 
 Figs.~\ref{fig:mf} and \ref{fig:bias}. On the left panels we can verify that 
 the $\Lambda$CDM simulations shown with red triangles are well described 
 by the HALOFIT fitting formula \cite{Smith:2002dz}, shown as a solid black line; for completeness 
the dotted line shows the linear power spectrum in this case. The results for the galileon simulations 
are shown as blue squares for $r_c=1665$ Mpc (top) and $r_c=1089$ Mpc (bottom). We also 
display in dashed lines the results from HALOFIT with different values of $\sigma_8$.

On the right panels we show percent differences  $\Delta P/P=(P^{r_c}/P^{\Lambda \text{CDM}}-1)$
for these cases. The triangles (squares) 
compare the $r_c=1665$ (1089) Mpc simulations to the $\Lambda$CDM simulation. 
The dashed lines compare the HALOFIT prediction at $\sigma_8=0.88$ (0.92) 
to the HALOFIT prediction at $\sigma_8=0.8$. 
  
Linear theory predicts a scale-independent deviation of $\sim 20\%$ and $\sim 32\%$
for $r_c=1665$ Mpc and 1089 Mpc respectively, relative to $\Lambda$CDM.  
This is in fact observed in the simulations on the largest linear scales.  
For mildly nonlinear scales, the HALOFIT prescription with higher 
$\sigma_8$ provides a close description of the departures seen in the galileon simulations 
relative to those of $\Lambda$CDM simulations. The departures are not well captured for scales that are 
more deeply inside the non-linear regime, due to the Vainshtein mechanism.  On these scales,
however, baryonic physical processes (not considered in our results) 
are in effect and represent an extra source of degeneracy.  

Therefore, our results from the power spectra in real space at $z=0$ indicate that 
on linear and mildly non-linear scales there is a degeneracy between galileons modified gravity 
and $\Lambda$CDM  with higher $\sigma_8$, similarly to that seen for the halo mass-function 
and linear bias.  Again, this degeneracy is redshift dependent and measurements of the 
power spectrum at distinct redshifts could in principle isolate features of galileon models. 
On deeply non-linear scales, the Vainshtein mechanism brings about unique features 
of galileon models at a single redshift, but processes due to the baryonic physics become 
more relevant, likely representing an even more important source of degeneracy, unless such 
complex processes are well characterized. 

\subsection{Modeling redshift space distortions \label{subsection:rsd_theory}}

The growth rate of structure  in the Universe may be determined through the observed anisotropy of the galaxy clustering in redshift space,
caused by the line of sight component of the galaxies peculiar velocities.
Redshift space effects alter the appearance of the clustering
of matter on all scales, and together with nonlinear evolution and bias, give rise to the measured anisotropic
power spectrum or correlation function  which is different from  the simple predictions of linear perturbation theory.
The comoving distance to a galaxy, ${\bf{s}}$,  
differs from its true distance, ${\bf{x}}$, due to its peculiar velocity, ${\bf{v}}({\bf{x}})$
(i.e. an additional velocity to the Hubble flow). The mapping from redshift space to real space is given by
\begin{eqnarray}
{\bf{s}} = {\bf{x}} + u_z {\bf \hat{z}},
\end{eqnarray}
where $u_z = {\bf{v}}\cdot {\bf \hat{z}}/(aH)$ and $H(a)$ is the Hubble parameter. This assumes that
 the distortions take place
 along the line of sight denoted by ${\bf \hat{z}}$. Note this is the plane parallel approximation which we adopt in this paper.

On large scales, coherent infall into overdense regions distorts  clustering statistics,
 causing the correlation function to appear squashed along the line of sight 
\cite[see][for a review of redshift space distortions]{1998ASSL..231..185H}.
For growing perturbations in the linear perturbation regime, the overall effect of redshift space distortions is 
to enhance the clustering amplitude.
This can be seen as an enhancement of the power spectrum in redshift space, 
$P_s({\bf{k}})$, compared to that in real space, $P_r(k)$.
This  effect was first analyzed by \cite{1987MNRAS.227....1K} in linear perturbation theory
and can be approximated by
\begin{equation}
\label{delta}
P_s(k,\mu) = P_r(k)(1+\mu^2 \beta)^2 ,
\end{equation}
where $\mu$ is the cosine of the angle between the wavevector, ${\bf{k}}$, and the line of sight.
The variable $\beta$ is
\begin{eqnarray}
\beta =  \frac{1}{b}\frac{{\rm{d ln}}D}{{\rm d ln}a} = \frac{f}{b} \, ,
\end{eqnarray}
where $f$ is referred to as the linear growth rate
 and  $b$ is the linear bias.
In this section we restrict the analysis to large scales and assume a constant bias $\hat{b} = \sqrt{\xi_{hh}(r)/\xi_m(r)}$ 
where $\xi_{hh(m)}$ is the two point correlation function for 
the halos (dark matter).
The \lq Kaiser formula\rq \,  (Eq.~\ref{delta}) relates the overdensity in redshift space to the corresponding value in real space and is the result of
 several approximations, e.g. that the velocity and density perturbations satisfy the linear continuity equation.
All of these assumptions are valid on scales that are well described by linear perturbation theory
and will break down on different scales as the density fluctuations grow 
\citep[see e.g.][for more details]{2004PhRvD..70h3007S,2011MNRAS.410.2081J, 2011ApJ...727L...9J}.

Rather than use the full 2D power spectrum, $P(k,\mu)$, it is common to decompose
the matter power spectrum in redshift space into multipole moments using Legendre polynomials, $L_l(\mu)$, \citep[see e.g.][]{1998ASSL..231..185H}
\begin{eqnarray}
P(k,\mu) = \sum_{l} P_l(k) L_l(\mu) \, ,
\end{eqnarray}
where the summation is over the order, $l$, of the multipole.
The anisotropy in $P({\bf{k}})$ is symmetric in $\mu$, as $P(k,\mu)=P(k,-\mu)$, so only even values of $l$ are summed over.
Each multipole moment is given by
\begin{eqnarray}
P^s_l(k) = \frac{2l+1}{2} \int_{-1}^{1} P(k,\mu) L_l(\mu) \rm{d}\mu \, ,
\end{eqnarray}
where the first two non-zero moments have Legendre polynomials, $L_0(\mu) = 1$ and  $L_2(\mu) = (3\mu^2 - 1)/2$.
Using the linear model in Eq.~\ref{delta}, the first multipole moment is given by
\begin{eqnarray}
 P_0(k)   
 &=& P_{m}(k)
(1 + \frac{2}{3}\beta + \frac{1}{5}\beta^2)  \,
\label{moments1}
\end{eqnarray}
where $P_{m}(k)$ denotes the real space matter power spectrum. Note we have 
omitted the superscript $s$ here
for clarity.

The corresponding equation in configuration space can be obtained by 
Fourier transforming Eq.~\ref{moments1} giving the corresponding relation between the monopole 
 of the correlation function in redshift space to
real space \citep{1998ASSL..231..185H}
\begin{eqnarray}
 \xi_0(s)   
 &=& 
\left(1 + \frac{2}{3}\beta + \frac{1}{5}\beta^2\right)\xi(r)   \,
\label{moments2}
\end{eqnarray}
where $\xi(r)$ is the real space correlation function. 
The above equations describe the boost in the clustering signal in redshift space on large scales where linear perturbation theory is valid.
To go beyond linear theory and deal with small scale velocities requires a model for the velocity field and all the density velocity correlations.
On small scales, randomized velocities associated with the motion of galaxies inside virialized structures reduce the power.
The dense central regions of galaxy clusters appear elongated along the line of sight in redshift space, which produces the  \lq fingers
of God\rq\
(FOG) effect seen in redshift survey plots \citep{1972MNRAS.156P...1J}.
This FOG effect can be described by convolving the correlation function
$\xi(s_{\perp},s_{||})$, where $s_{||}$ is the distance separation along the line of sight and $s_{\perp}$ is the perpendicular separation,
 with the distribution function of random pairwise velocities, $f(u)$ \citep{1976Ap&SS..45....3P},
\begin{eqnarray}
\xi(s_{\perp},s_{||}) &=& \int_{-\infty}^{\infty} {\rm d}u f(u) \xi(s_{\perp},s_{||}-u) \, ,
\end{eqnarray}
where $f(u)$ can have an exponential or a Gaussian form such as
\begin{eqnarray}
f(u) &=& \frac{1}{\sqrt{2\pi\sigma^2_v}}{\rm exp}\left(-\frac{u^2}{2\sigma^2_v} \right) \, ,
\end{eqnarray}
where $\sigma_v$ is the pairwise peculiar velocity dispersion.
This model has been used to fit to results from both simulations and observations 
\citep[see, for example][]{2004PhRvD..70h3007S,2009MNRAS.393..297P,2011MNRAS.410.2081J,2008Natur.451..541G, 2011MNRAS.415.2876B, 2012MNRAS.423.3430B}.
 Recently there have been
 many models which  improve on this description of redshift space distortions in the nonlinear 
regime \citep{2004PhRvD..70h3007S, 2008PhRvD..78h3519M, 2010PhRvD..82f3522T, 2011MNRAS.417.1913R, 2011JCAP...11..039S}.
In this work our main concern is to quantify the relative difference in the correlation function measured from the simulations 
in redshift and real space in $\Lambda$CDM compared to 
galileon models. We restrict our study to large scales and shall compare the measured ratio between the redshift 
and real space correlation function with the linear perturbation theory predictions in each cosmology.

\begin{figure}[htbp] 
\centering
\includegraphics[width=0.45\textwidth]{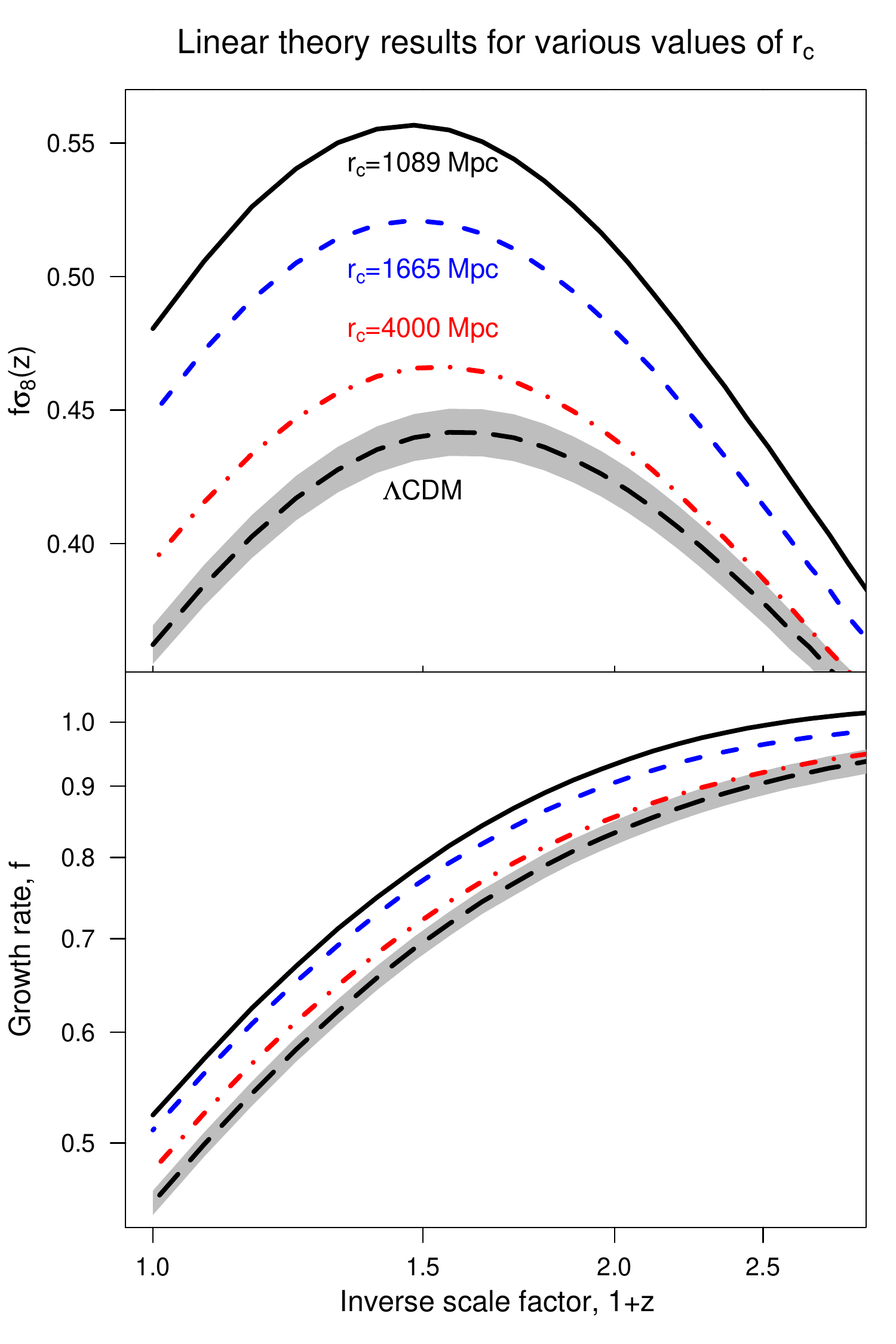}
   \caption{The linear growth rate -- both weighted by $\sigma_8(z)$ and alone --
    as a function of redshift, $z$, for $\Lambda$CDM, the $r_c = 1089$~Mpc and  $r_c = 1665$~Mpc
galileon models,  plus a comparison model with $r_c=4000$~Mpc,
 are shown as a  black dashed line, a solid black line, a dashed blue line, and a dash-dotted red line, respectively. 
 The grey band around the $\Lambda$CDM line represents the expected precision of future surveys, which hope to achieve
 $\pm 2$\% accuracy in their measurement of the growth rate. }
   \label{fig:growthrate}
\end{figure}

\subsection{Redshift space distortions from galileons \label{subsection:rsd_results}}

Measuring the anisotropic distortions in the galaxy clustering pattern in redshift space
constrains the parameter $\beta = f/b$, where
$b$ is the galaxy bias factor and $f$ is the logarithmic derivative of the linear growth rate of structure, which
is scale independent in the case of general relativity. 
In the lower panel of Fig.~\ref{fig:growthrate} we show the linear growth rate as a 
function of redshift for  $\Lambda$CDM and the $r_c = 1089$ Mpc and the $r_c = 1665$ Mpc
galileon models as a dashed black line, a solid black line  and a dashed blue line respectively. A model with
$r_c = 4000$ Mpc is also plotted as dot-dashed red line for comparison. In the upper panel we plot the linear growth rate weighted by $\sigma_8(z)$ for each model.
The grey shaded region around the $\Lambda$CDM result represents the expected precision of a DETF \cite{2006astro.ph..9591A} 
Stage IV galaxy redshift survey such as the ESA's
EUCLID mission \cite{2011arXiv1110.3193L},  WFIRST \cite{2012arXiv1208.4012G} or the ground-based 
dark energy experiment, BigBOSS \citep{2009arXiv0904.0468S}, which aim to 
achieve $\sim 2\%$ accuracy in their measurement
of the growth rate. 
The relative difference in the linear growth rate between the $r_c = 1089$ Mpc model and $\Lambda$CDM varies from 16\% at $z=0$ to 12\% at $z=1$,
while for $r_c = 1665$ Mpc the difference is 12\% and 8\% at redshifts $z=0$ and $z=1$ respectively.
It is clear that if future galaxy redshift surveys can  measure the growth rate to within 2\%, they 
will be able to place significant constraints on currently allowed galileon modified gravity models.

\begin{figure*}[htbp] 
\centering
   \includegraphics[bb= 69 364 515 616,width=0.85 \textwidth]{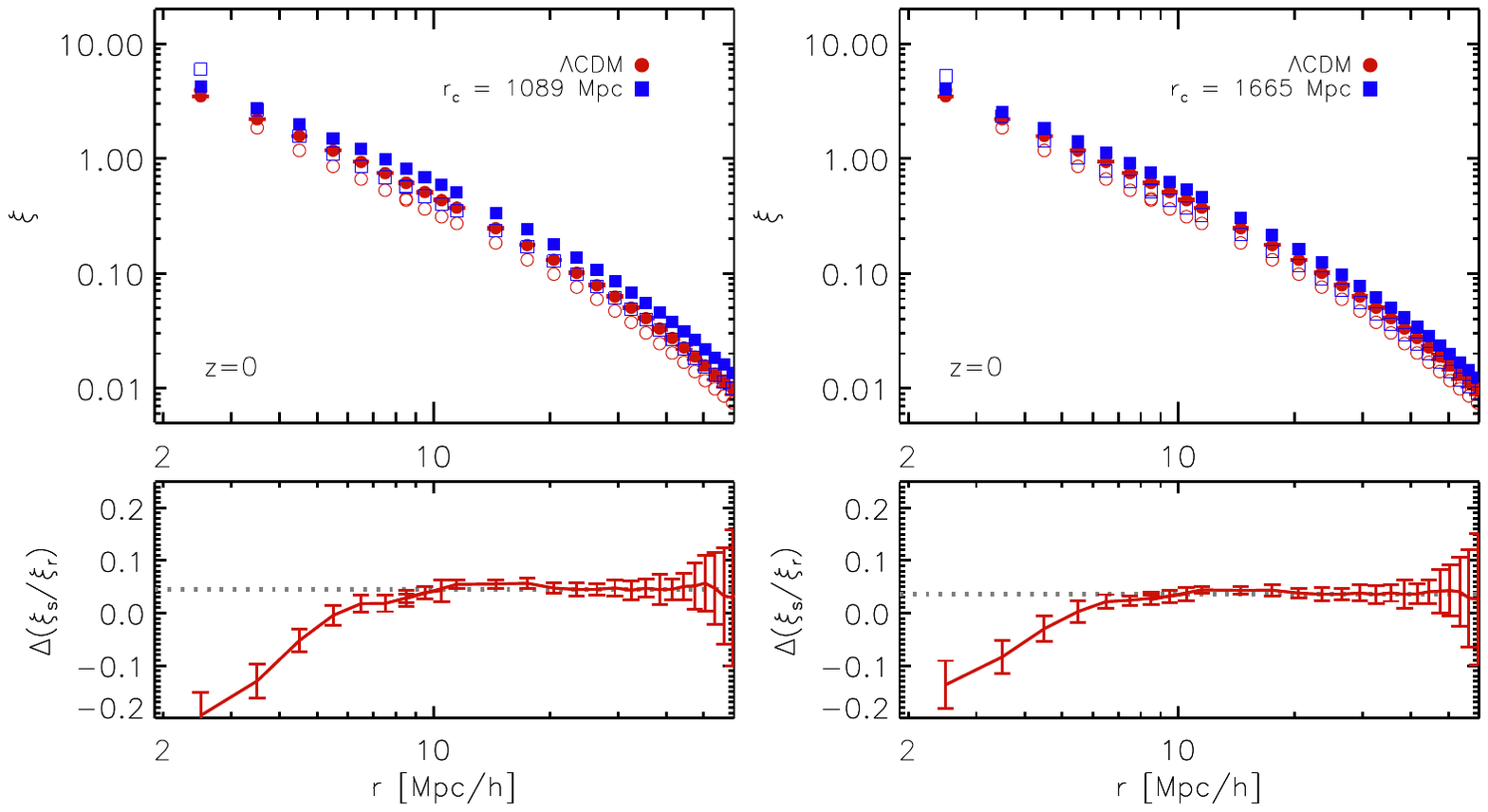}
   \caption{Upper panels: The dark matter two point correlation function measured  in real and redshift space at $z=0$ are shown as red circles and blue squares for
$\Lambda$CDM and the $r_c = 1089$ Mpc ($r_c = 1665$ Mpc) model in the left (right) panel. 
Open symbols denote the correlation function measured in real space while closed symbols represent the correlation function measured in redshift space.
Lower panels: The relative difference in the ratio of the correlation function measured in redshift space to that in real space,
 in the 
$r_c = 1089$ Mpc ($r_c = 1665$ Mpc) model compared to $\Lambda$CDM is shown in the left (right) panel.
The dotted grey line in both the right and left panel shows the Kaiser linear theory prediction for this ratio using the appropriate growth rate for each model.
}
   \label{fig:z0}
\end{figure*}

\begin{figure*}[htbp] 
   \centering
  \includegraphics[bb= 59 361 510 625,width=0.85 \textwidth]{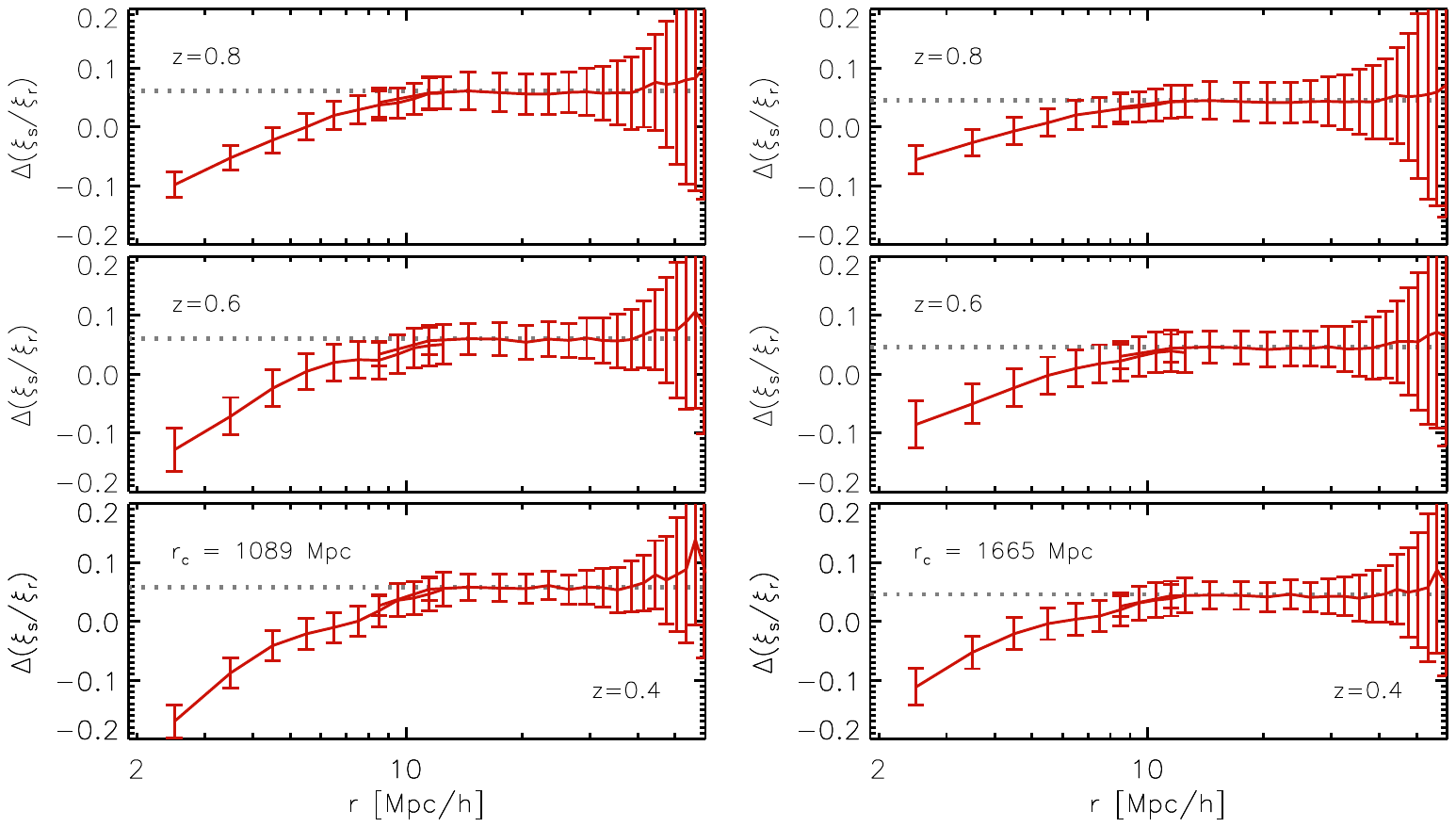}
   \caption{The relative difference in the ratio of the correlation function measured in redshift space to that in real space,
 in the
$r_c = 1089$ Mpc ($r_c = 1665$ Mpc) model compared to $\Lambda$CDM is shown in the left (right) panel. The upper, middle and lower
panels show this ratio measured from the simulatons at $z=0.8$, $z=0.6$ and $z=0.4$ respectively with Jackknife errors on the mean.
The dotted grey line in both the right and left panels shows the Kaiser linear theory prediction for this ratio using the appropriate growth rate for each model
at the redshift indicated by the legend.
}
   \label{fig:z04_z08}
\end{figure*}

In Fig.~\ref{fig:z0} we show the two point correlation function in real and redshift space measured 
from the simulations at $z=0$ for $\Lambda$CDM and the $r_c = 1089$ Mpc ($r_c = 1665$ Mpc) model in the left (right) panels.
In the upper panels open symbols represent the correlation function in real space while closed symbols are used for the redshift space function $\xi(s)$.
It is clear from this figure that both the real and redshift space correlation function amplitude are increased in the modified gravity models 
compared to $\Lambda$CDM. We plot the relative difference in the 
ratio of the correlation function measured in redshift space to that in real space,
 in the
$r_c = 1089$ Mpc ($r_c = 1665$ Mpc) model compared to $\Lambda$CDM in the lower left (right) panel. 
Here $\Delta(\xi_s/\xi_r) = (\xi_s/\xi_r)_{r_c}/(\xi_s/\xi_r)_{\tiny \mbox{$\Lambda$CDM}} -1 $ 
where $(\xi_s/\xi_r)_{r_c}$ is the ratio of the monopole of the redshift to real space correlation function for the modified gravity model.

\begin{figure}[htbp] 
   \centering
  \includegraphics[bb= 70 364 286 575,width=0.45 \textwidth]{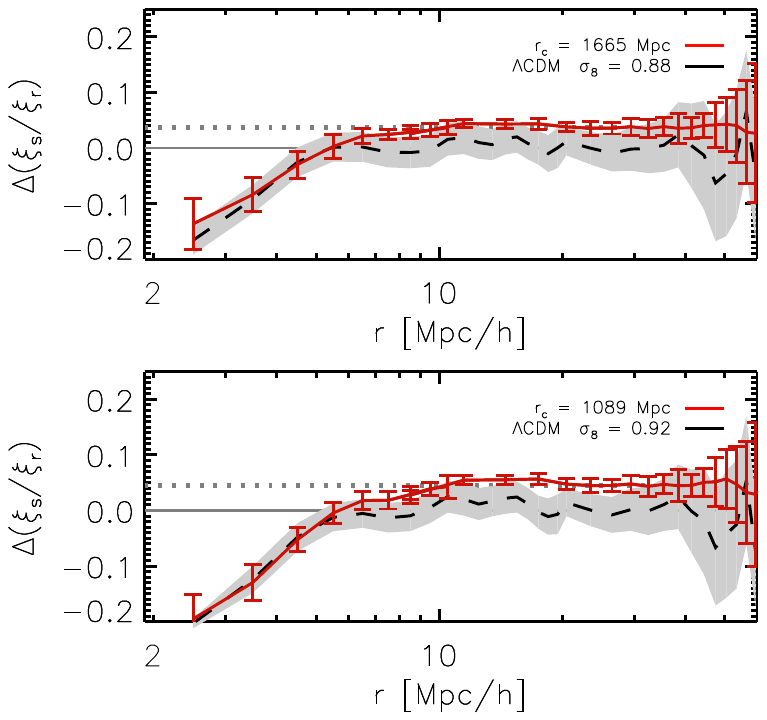}
   \caption{The relative difference in the ratio of the correlation function measured in redshift space to that in real space,
 in the
$r_c = 1089$ Mpc ($r_c = 1665$ Mpc) model compared to $\Lambda$CDM is shown in the top (bottom) panel as a red solid line with error bars 
as in Fig.~\ref{fig:z0}. The difference in the ratio $\xi_s/\xi_r$ measured from a $\Lambda$CDM simulation with $\sigma_8 =0.92$ 
($\sigma_8=0.88$) to a $\Lambda$CDM simulation with $\sigma_8 =0.8$ is shown as a black dashed line in the left (right) panel. The grey shaded 
regions represent the jackknife errors on the mean.   
}
   \label{fig:new_sigma}
\end{figure}

From Fig.~\ref{fig:z0} we find an increase in the clustering signal in redshift space on large scales in the modified gravity model compared to $\Lambda$CDM
and an increase in the small scale damping due to incoherent random velocities. This increased clustering signal at $r > 10$ Mpc$/h$ in the 
modified gravity models is due to increased bulk flows on large scales. On small scales  the enhanced forces in the modified gravity model create a 
larger velocity dispersion which gives rise to increase damping compared to $\Lambda$CDM on scales $r < 10$ Mpc$/h$. 
These results agree with a similar study of redshift space distortions in $f(R)$ modified gravity carried out by \cite{Jennings:2012pt}.
The Kaiser linear theory prediction for this ratio using the appropriate growth rate for each model is shown in the lower panels in Fig.~\ref{fig:z0} as a dotted grey line. The measured difference in the 
redshift to real space ratio between $\Lambda$CDM and galileon models agree with linear theory 
predictions on scales $r > 15$ Mpc$/h$ for the $r_c = 1089$ Mpc model and $r > 8$ Mpc$/h$ for $r_c = 1665$ Mpc.

\begin{figure*}[htbp] 
   \centering
  \includegraphics[bb= 60 362 511 624,width=0.85 \textwidth]{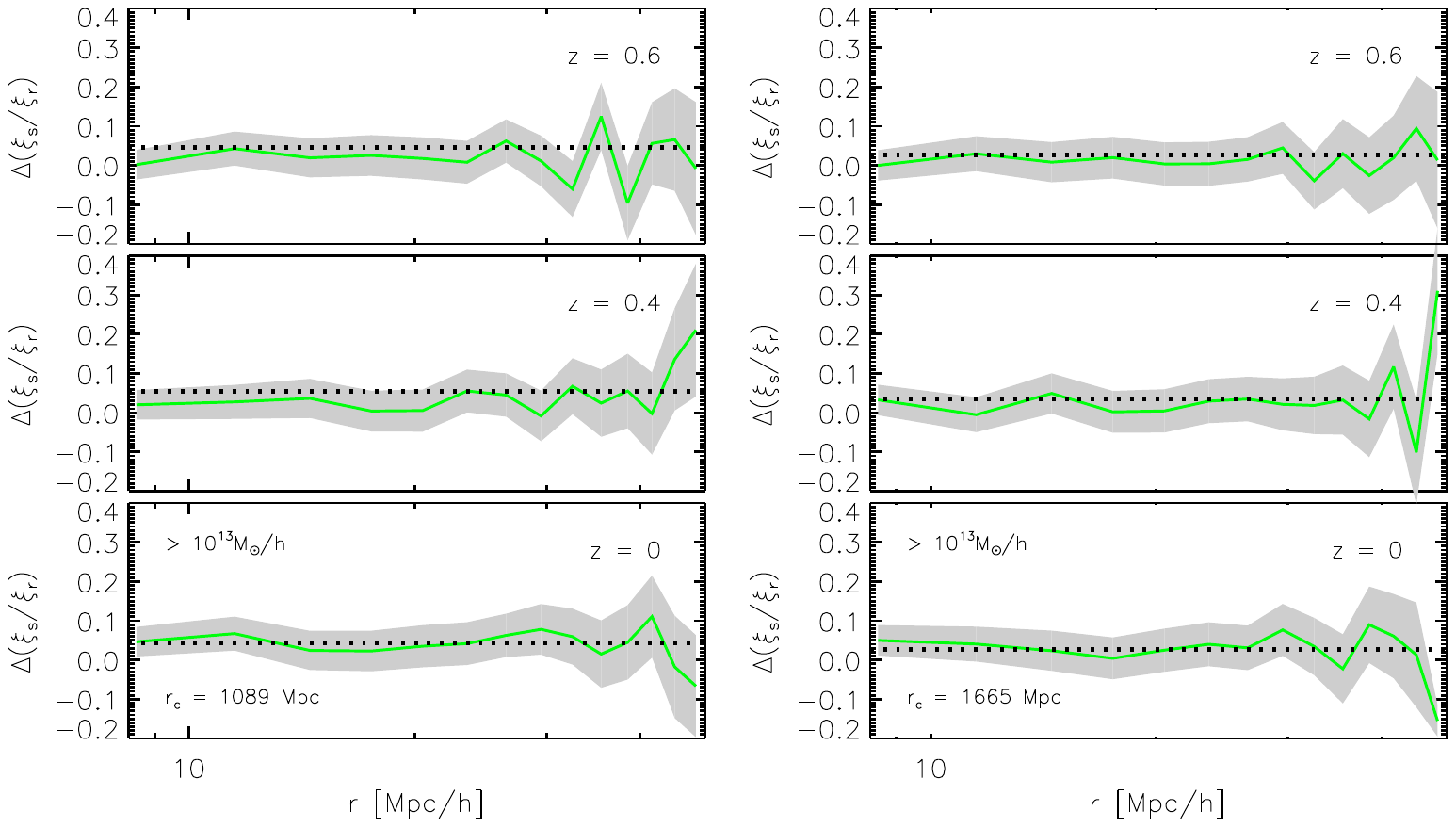}
   \caption{The ratio of $\xi_s/\xi_r$ measured using all halos with masses $>10^{13}M_{\odot}/h$ 
in the $r_c = 1089$ Mpc ($r_c = 1665$ Mpc) model  compared to $\Lambda$CDM 
are shown in the left (right) panels as a green solid line. The results at $z=0$, $z=0.4$ and $z=0.6$  are shown 
in the bottom, middle and top panels respectively. The grey shaded region show the Jackknife errors on the mean.
}
   \label{fig:halos}
\end{figure*}

Both the upper and lower panels in Fig.~\ref{fig:z0} combine measurements from the $L_{\tiny \mbox{box}} =400$ Mpc$/h$ and $L_{\tiny \mbox{box}} =256$ Mpc$/h$ 
simulations. These measurements represent the average over 8 realizations and the errors plotted represent the scatter amongst these 8 different 
simulations for each cosmological model. We plot the ratio as in the lower panels in Fig.~\ref{fig:z0} to remove any sample variance in the measurements 
which arise from sampling a finite number of large scale modes in a finite simulation volume.

In Fig.~\ref{fig:z04_z08} we plot the relative difference, as in the lower panel in Fig.~\ref{fig:z0},
 for the $r_c = 1089$ Mpc (left) and $r_c = 1665$ Mpc (right) model at $z=0.4$, $z=0.6$ and $z=0.8$ in the lower, middle and upper panels respectively.
The relative difference in the ratios increases with redshift from e.g. 4\% at $z=0$ to 6\% at $z=0.8$ for the $r_c = 1089$ Mpc model and agrees with the linear
perturbation theory predictions (grey dotted line) on larger scales compared to $z=0$. 
The enhanced forces due to the galileon scalar field decrease with increasing redshift and as a result the relative increase in the 
small scale damping decreases.
The differences between the modified gravity models considered here and standard gravity
 decrease with increasing redshift and so it may be counterintuitive that the ratios shown in 
Fig.~\ref{fig:z04_z08} increase with redshift on large scales. The reason for this can be found by examining the linear perturbation 
theory function for the ratio $\xi_s/\xi_r = 1 +2/3f+1/5f^2$ for the dark matter. For a given cosmological model, this ratio 
increases with increasing $f$ (increasing redshift). Even though
the relative difference in $f$ between a modified gravity cosmology and $\Lambda$CDM decreases from $z=0$ to $z=0.8$, the 
relative increase in the ratio will be larger at $z=0.8$ compared to $z=0$, as can been seen from Fig.~\ref{fig:z04_z08}.
The errors plotted here represent the Jackknife errors on the mean as outlined in Appendix \ref{section:xi}.

From Figs.~\ref{fig:mf} and \ref{fig:bias}, it is clear that both the halo mass function and linear bias, calculated using the Tinker fitting formulae 
\cite{Tinker:2008ff, Tinker:2010my} can match the measurements from the $r_c = 1089$ Mpc and 
the $r_c = 1665$ Mpc  simulations if we assume a $\Lambda$CDM cosmology with a higher value for $\sigma_8$. 
Moreover, from Fig.~\ref{fig:ps}, the non-linear real-space power spectrum calculated using the HALOFIT fitting formula 
\cite{Smith:2002dz} can also match the modified gravity at mildly non-linear scales.
We test if this degeneracy also occurs in the measurements of the correlation function in 
redshift space by running two additional $\Lambda$CDM  simulations using the same $\Omega_{\rm m}$ but with $\sigma_8=0.92$ and $\sigma_8=0.88$ at $z=0$. 
In Fig.~\ref{fig:new_sigma} we plot the same ratio of $\xi_s/\xi_r$ for the
$r_c = 1089$ Mpc ($r_c = 1665$ Mpc) model compared to $\Lambda$CDM in the top (bottom) panel as a red solid line with error bars, as shown in 
Fig.~\ref{fig:z0}.
The relative difference in $\xi_s/\xi_r$ measured from a $\Lambda$CDM simulation with $\sigma_8 =0.92$
($\sigma_8=0.88$) compared to a $\Lambda$CDM simulation with $\sigma_8 =0.8$ is shown as a black dashed line in the top (bottom) panel. The grey shaded
regions represent the jackknife errors on the mean. 

The difference in the monopole to real space correlation function 
between $\Lambda$CDM cosmologies which have different power spectrum amplitudes at $z=0$ has a  different signature on large scales compared to the measurements 
 from our modified gravity simulations.  The ratio on large scales is consistent with unity 
and is distinguishable from the relative increase in this ratio measured in the modified gravity simulations compared to $\Lambda$CDM, 
while on small scales we measure a similiar increase in the damping signal in the $\Lambda$CDM 
simulation with $\sigma_8=0.92$ ($\sigma_8=0.88$) compared to the simulation with $\sigma_8=0.8$. In linear perturbation theory the velocity, $v \propto f\sigma_8$ and in 
this case
the growth rate, $f$, is the same for all the $\Lambda$CDM simulations. This implies that the velocities for the $\sigma_8=0.88 \;(0.92)$ simulation will be higher then in the 
$\sigma_8=0.8$  case and explains the decrease in the ratio on small scales shown as a black dashed line in Fig.~\ref{fig:new_sigma}. 
These results agree with similar studies of the correlation function in real and redshift 
space for $\Lambda$CDM cosmologies varying the parameter combination 
$(\Omega_{\rm m}, \sigma_8)$ which were carried by \cite{2002ApJ...575..617Z,2006MNRAS.368...85T}.

In Fig.~\ref{fig:halos} we plot the ratio of $\xi_s/\xi_r$ measured using all halos with masses $>10^{13}M_{\odot}/h$
in the $r_c = 1089$ Mpc ($r_c = 1665$ Mpc) simulation  compared to $\Lambda$CDM
in the left (right) panels as a green solid line. The mean value plotted in this figure is from two simulations and the gray shaded region represents the Jackknife 
errors on this mean. 
The resolution of our simulations limits us to using only the 400 Mpc$/h$ simulation boxes for this measurement and so we cannot probe the
difference in the nonlinear damping signals for these halo mass ranges at $r<10$ Mpc$/h$.

The dotted black line represents the predictions of Eq.~\ref{moments2} where $\beta = f/b$ and we use the linear theory value for the growth rate
in each cosmology at each redshift. We take the linear bias to be the best fit value for the ratio  
$b(M) = \sqrt{\xi_{ hh}(r,M)/\xi_{ m}(r)}$ over 
the range $r=20-50$Mpc$/h$, where $\xi_{ hh}(r,M)$  
is the halo correlation function in real space and $\xi_{ m }(r)$ is the dark matter correlation function.
We find that the bias for $\Lambda$CDM  halos in the mass range $>10^{13} M_{\odot}/h$ varies from $b\approx2$ at $z=0$ to $b\approx3.9$ at $z=0.8$ compared to the 
$r_c = 1089$ Mpc model which has $b=1.5$ ($b=2.9$) at $z=0$($z=0.8$). The measured relative ratios in Fig.~\ref{fig:halos} are consistent with 
the linear perturbation theory prediction at all redshifts within the error bars with a 4\% (5\%) difference between the $r_c = 1089$ Mpc model and 
$\Lambda$CDM at $z=0$ ($z=0.8$). 
A similar analysis for coupled dark energy cosmologies was recently carried out by \cite{2012MNRAS.420.2377M} where they also examined the two point 
correlation function of halos in redshift and real space at several redshifts.

\section{Conclusions}
We have studied linear and non-linear structure formation in a class of modified gravity models with a $\Lambda$CDM-like expansion
history and a galileon scalar field that is screened in regions of high density via the Vainshtein mechanism. Our primary results have been derived from a large
suite of N-body simulations of this model, where the effects of the extra scalar field have been captured through direct numerical solution
of its non-linear equations of motion. We have found, as expected, that large scale structure is enhanced in this model relative to ordinary
general relativity. This enhancement appears through a significant increase in the number of large halos formed in this model relative
to standard gravity, with the enhancement matching that anticipated from linear theory: the
modified gravity models lead to a late-time power spectrum normalization of $\sigma_8 =0.92$ ($0.88$) 
for the values of $r_c$ that we studied, $r_c=1089$ (1665) Mpc. Hence, for a single redshift, the effect of the galileon field is degenerate
with an increase in the primordial amplitude of fluctuations. However, as we emphasized in Fig.~\ref{fig:lineartheory}, the apparent $z=0$ 
normalization of the power spectrum is itself a function of redshift in this theory, so measurements of the halo mass function
at different redshifts would break this degeneracy.  This degeneracy also appears in the real space power spectrum even 
at mildly non-linear scales. On deeply non-linear scales the Vainshtein screening provides a unique feature of galileon models, but 
is likely degenerate with the much more complex baryonic physics that becomes more relevant on these scales. 

Next, we examined the impact of redshift space distortions on the measured two point correlation function in our simulations. We find 
 deviations in the ratio of the redshift to real space correlation function $\xi_s/\xi_r$ in the modified gravity models that are clearly
distinguishable from standard gravity. On large scales the redshift space correlation function is sensitive to the growth rate, which for the
modified gravity models we consider here,  can be dramatically different from the growth rate in standard gravity. 
For the values of $r_c$ that we studied, we found deviations in the dark matter clustering signal on large length scales at the level of 10\% 
and a difference of 4-5\% at $z=0-0.8$ for halos $>10^{13} M_{\odot}/h$ which is potentially
large enough to be seen or ruled out by  future galaxy redshift  surveys. As we show in Fig.~\ref{fig:new_sigma}, the
redshift space distortion signal on large scales
cannot be mimicked by increasing the initial amplitude of fluctuations and using standard gravity. Going beyond
what we can model in linear theory, we also find that the enhanced gravitational force
gives rise to a diminution of the clustering signal on small ($< 9 $ Mpc$/h$) length scales beyond that seen in GR,
 as the enhanced non-linearities more efficiently wash out the correlated motions found on large length scales.
 
Let us compare our results to those found in $f(R)$ gravity. In contrast with our findings, the halo mass functions for $f(R)$ models are not well approximated by simple changes in $\sigma_8$, even for a single redshift. Furthermore,  since the extra scalar field that appears in $f(R)$ is always massive,
it necessarily has a range limited by Yukawa suppression. Hence, $f(R)$ models cannot generate the long-range deviations from
GR that are present in the redshift space correlations we have studied in galileon models. This is an important illustration of how $f(R)$ gravity and galileon
models are quite distinct in their phenomenology.

Despite the presence of Vainshtein screening in regions of high density, our findings suggest that
 linear perturbation theory is a good guide to understanding how large scale structure is modified
in models with a galileon scalar field, at least for the values of $r_c$ we have studied. If, as data improve, the limits on $r_c$ are pushed upwards
towards today's Hubble scale, the methods we employ will have to improve accordingly to make further progress. On the theoretical side,
it will be necessary to understand how to improve the phenomenological model described in  \S \ref{pheno} to represent better the interplay
between the cosmological and local excitations of the scalar mode of the graviton. On the computational side, our simulation methods will have to improve
in order to capture a greater range of scales so that we can find and quantify the scalar's effects as they become more subtle, as they will
with larger $r_c$.  Since $r_c \sim c/H_0$ is the value we expect if today's cosmological acceleration is generated by a massive graviton, 
we will only be able to use the growth of structure to  constrain the theory definitively once all of these improvements are made. 

\emph{Note added:} While this manuscript was in the final stages of preparation, Ref. \cite{Li:2013nua} appeared, describing a new code that can
solve galileon-type equations on an adaptively refined mesh. However, they study the model of self-accelerating DGP, 
which has a different expansion history from $\Lambda$CDM and a repulsive, rather than an attractive, extra scalar force. Hence their results
are physically distinct from the phenomenological model studied in this paper.

\acknowledgements
We thank W. Hu, A. Klypin, A. Kravtsov, and R. Wechsler for helpful discussions, and J. Khoury for collaboration in an earlier stage of this work. Some of the numerical simulations reported here were performed on a cluster supported in part by the Kavli Institute for Cosmological Physics at the University of Chicago through grants NSF PHY-0114422 and NSF PHY-0551142 and an endowment from the Kavli Foundation and its founder Fred Kavli. 
We also acknowledge  resources provided by the University of Chicago Research Computing Center. MW was supported by U.S. Dept. of Energy contract DE-FG02-90ER-40560.
EJ acknowledges the support of a grant from the Simons Foundation, award number 184549. 
ML is supported by FAPESP and CNPq. 
\appendix

\section{Numerical details} \label{numerics}

For our N-body simulations, we improved the code first reported on in \cite{Khoury:2009tk},
solving a very similar set of equations as those studied in \cite{Schmidt:2009sg}. The code is written in FORTRAN
and is largely based on the publicly available PM-Code \cite{Klypin:1997sk}, a particle-mesh N-body code that
employs fast fourier transforms to solve the Newtonian Poisson equation and cloud-in-cell grid assignment to interpolate
discrete particle positions onto the density grid. The primary addition we have made to the public code beyond
those reported in \cite{Khoury:2009tk} is the inclusion of a multigrid relaxation subroutine for solving the non-linear
equation for the extra scalar field, Eqs.~\ref{sys}. This subroutine is heavily adapted from \cite{Press:1996:NRF:232468}.
We have also included threaded parallelization through OpenMP.
The numerical method we utilize is substantially identical to that described in Appendix A of \cite{Schmidt:2009sg}.  
Results using this version of the code first appeared in \cite{Wyman:2010jp}.
The discretization method we use for the derivatives that appear in Eq.~\ref{sys} is a standard one; e.g., 
for a field $\phi$ at the grid location $\{i,j,k\}$ on an $x,y,z$ grid, we would have
\begin{align}
\nabla_x \nabla_x \phi_{i,j,k} = h^{-2} & \(\phi_{i+1,j,k} + \phi_{i-1,j,k} - 2 \phi_{i,j,k}\) \\
\nabla_x \nabla_y \phi_{i,j,k} = \frac{1}{4} h^{-2} & \( \phi_{i+1,j+1,k} - \phi_{i+1,j-1,k}  \right . \nonumber \\
& \left. - \phi_{i-1,j+1,k} + \phi_{i-1,j-1,k}\)
\end{align}
and $h=2^n$, with $n$ representing the level of refinement in the multigrid relaxation.

\section{Two point clustering statistics in real and redshift space \label{section:xi}}

Calculating the two point correlation 
function for $N$ particles by direct 
pair counting requires $N^2$ operations. Considering the 
large number of particles used in the simulation we make use of an estimator introduced  by
\cite{2002MNRAS.333..443B,2008MNRAS.390.1470S}. In this approach a density 
field is constructed on $N_{\tiny \mbox{grid}}$ cells and the correlation function is then calculated as
\begin{eqnarray}
\hat{\xi}(|r_{ij}|) &=& \frac{1}{N_{\rm{run}}N_{p}(|r_{ij}|)} \sum_{k=1}^{\rm{N_{run}}}\sum_{ij} (\delta(r_i) \delta(r_j))_k
\label{xi}
\end{eqnarray}
where $\delta(r_i)) = (n(r_i) - <\bar{n}>)/<\bar{n}>$ is the density fluctuation in the
$i^{\tiny \mbox{th}}$ bin of the grid. The sum extends over all $N_{p}$ pairs
separated by distances between $r - \Delta r/2$ and $r+\Delta r/2$. We also sum over 
eight realisations for each simulation of a particular cosmological model, $N_{\tiny \mbox{run}} =8$.
This procedure scales as $N_{\tiny \mbox{grid}}^2$ and requires fewer operations then direct pair counting as usually $N{\tiny \mbox{grid}} \ll N$.

This approach is a far more efficient method to measure the correlation function than direct pair counting and has been shown to be 
 extremely accurate at 
reproducing the full two point function for a range of grid sizes \citep[see Figure 5 in][]{2008MNRAS.390.1470S}.
\cite{2009PhRvD..80l3503T} and \cite{2011PhRvD..84d3501S} also used a grid based calculation with FFT to measure the correlation function and found that this method accurately 
reproduces the $\xi(r)$ found from direct pair counting.
Using a lower resolution simulation, we have verified that the using the estimator in Eq.~\ref{xi} reproduces the correlation function measured
 using the standard 
\cite{1993ApJ...412...64L} method.

This grid based method
limits the accuracy of our measurements to
scales larger than a few grid cells, $r>L_{\tiny \mbox{box}}/N_{\tiny \mbox{grid}}$.
For the $L_{\tiny \mbox{box}} = 400$ Mpc$/h$ simulation we use a $N_{\tiny \mbox{grid}} = 200^3$ grid while for the $L_{\tiny \mbox{box}} = 256$ Mpc$/h$ 
simulation we use $N_{\tiny \mbox{grid}} = 256^3$. Errors on the $z=0$ correlation function represent the scatter amongst 8 realisations of the same cosmology 
where different random number seeds where used to generate the initial conditions for the simulations. The errors on 
the measurements at $z>0$ were obtained by Jackkife sampling 
from a single simulation by dividing the simulation volume into $N_{\tiny \mbox{sub}}=8$ equal subvolumes 
and then systematically omitting one subvolume at a time in order to calculate the 
correlation function on the remaining  $N_{\tiny \mbox{sub}}-1$ volume \citep[see][for more details of this method]{nbg2009}.
The redshift space correlation function is obtained from the simulations after averaging over the
$\xi(s)$ obtained by treating the $x, y$ and $z$ directions  in turn as the lines of sight.

\bibliography{HaloBib}

\end{document}